\documentclass[12pt,a4paper]{article}

\usepackage{epsf}
\usepackage{latexsym,amssymb,euscript}
\usepackage[dvips]{graphicx}
\usepackage{amsmath}

\usepackage[italian,english]{babel}
\usepackage{indentfirst}
\usepackage[T1]{fontenc}
\usepackage{graphicx,xcolor}
\usepackage[format=hang,font=small]{caption}
\usepackage{subfig}
\usepackage{amsmath}
\usepackage{amsfonts}
\usepackage{lscape}
\usepackage{amsmath}
\usepackage{amssymb}
\usepackage{hyperref}
\usepackage{geometry}
\usepackage{slashed}

\usepackage{epsf}
\usepackage{latexsym,amssymb,euscript}

\newcommand{\beq}[1]{
\begin{equation}\label{#1}}
\newcommand{\eeq}{\end{equation}}
\newcommand{\bea}[1]{
\begin{eqnarray}\label{#1}}
\newcommand{\eea}{\end{eqnarray}}

\makeatletter
\def\appendix{\par\clearpage
  \setcounter{section}{0}
  \setcounter{subsection}{0}
  \@addtoreset{equation}{section}
  \def\@sectname{Appendix~}
  \def\theequation{\thesection.\arabic{equation}}
  \def\theequation{\thesection.\arabic{equation}}
  \def\thesection{\Alph{section}}}
\makeatother

\begin{document}
\begin{titlepage}

\begin{center}
{\LARGE \bf On the BLM optimal renormalization scale 

setting for semihard processes}
\end{center}

\vskip 0.5cm

\centerline{F.~Caporale$^{1\ast}$, D.Yu.~Ivanov$^{2\P}$, B.~Murdaca$^{3\dagger}$ 
and A.~Papa$^{3\dagger}$}

\vskip .6cm

\centerline{${}^1$ {\sl Instituto de F\'{\i}sica Te\'orica, Universidad 
Aut\'onoma de Madrid,}}
\centerline{\sl E-28049 Madrid, Spain}

\vskip .2cm

\centerline{${}^2$ {\sl Sobolev Institute of Mathematics and
Novosibirsk State University,}}
\centerline{\sl RU-630090 Novosibirsk, Russia}

\vskip .2cm

\centerline{${}^3$ {\sl Dipartimento di Fisica, Universit\`a della Calabria,}}
\centerline{\sl and Istituto Nazionale di Fisica Nucleare, Gruppo collegato di
Cosenza,}
\centerline{\sl I-87036 Arcavacata di Rende, Cosenza, Italy}

\vskip 2cm

\begin{abstract}
The BFKL approach for the investigation of semihard processes is plagued
by large next-to-leading corrections, both in the kernel of the universal
BFKL Green's function and in the process-dependent impact factors, as well
as by large uncertainties in the renormalization scale setting. All that
calls for some optimization procedure of the perturbative series. In this 
respect, one of the most common methods is the Brodsky-Lepage-Mackenzie (BLM) 
one, that eliminates the renormalization scale ambiguity by absorbing the 
non-conformal $\beta_0$-terms into the running coupling.
In this paper, we apply BLM scale setting procedure directly to the 
amplitudes (cross sections) of several semihard processes. We show that,
due to the presence of $\beta_0$-terms in the next-to-leading expressions
for the impact factors, the optimal renormalization scale is not universal, 
but depends both on the energy and on the type of process in question.
\end{abstract}

\vskip .5cm

$
\begin{array}{ll}
^{\ast}\mbox{{\it e-mail address:}} & \mbox{francesco.caporale@uam.es}\\
^{\P}\mbox{{\it e-mail address:}} & \mbox{d-ivanov@math.nsc.ru}\\
^{\dagger}\mbox{{\it e-mail address:}} &
\mbox{beatrice.murdaca, alessandro.papa \ @fis.unical.it}\\
\end{array}
$

\end{titlepage}

\section{Introduction}

We discuss the application of the BFKL method~\cite{BFKL} to the description of 
semihard processes, i.e. hard processes in the kinematic region where the 
energy variable $s$ is substantially larger than the hard scale $Q^2$, 
$s\gg Q^2\gg \Lambda_{\rm QCD}^2$, with $Q$ the typical transverse momentum and 
$\Lambda_{\rm QCD}$ the QCD scale. This approach allows to resum systematically 
to all orders of perturbation series the terms enhanced by leading 
$\alpha_s^n \ln^n(s/Q^2)$ (LLA) and first subleading $\alpha_s^{n+1}\ln^n(s/Q^2)$
(NLA) logarithms of the energy. 

In the BFKL approach, relevant physical observables are expressed as a 
convolution of two impact factors with the Green's function of the BFKL 
equation. The Green's function is determined through the BFKL equation and is
process-inde\-pen\-dent. The next-to-leading order (NLO) kernel of the BFKL 
equation for singlet color representation in the $t$-channel and forward 
scattering, relevant for the determination of the NLA total cross section, has 
been achieved in Refs.~\cite{NLA-kernel}, after the long program of calculation
of the NLO corrections~\cite{NLA-corrections} (for a review, see
Ref.~\cite{news}). The other essential ingredient are impact factors, which 
are not universal and must be calculated process by process. Indeed, only a 
few of them are known with NLO accuracy.

Both the impact factors and the BFKL kernel receive large NLO corrections in 
the $\overline{\rm{MS}}$ renormalization scheme. The practical application
of this approach to physical processes encounters therefore serious 
difficulties, due not only to the large NLO corrections, but also to 
big renormalization scale setting uncertainties, thus calling for some 
optimization procedure of the QCD perturbative series. 

In this paper we focus on the widely-used Brodsky-Lepage-Mackenzie (BLM) 
approach~\cite{BLM} to face with this problem, which relies on the removal
of the renormalization scale ambiguity by absorbing the non-conformal 
$\beta_0$-terms into the running coupling. It is known that after BLM scale 
setting, the QCD perturbative convergence can be greatly improved due to the 
elimination of renormalon terms in the perturbative QCD series. Moreover, with 
the BLM scale setting, the BFKL Pomeron intercept has a weak dependence on the 
virtuality of the Reggeized gluon~\cite{Brodsky:1998kn,Brodsky:2002ka}.

We apply the BLM scale setting procedure directly to the amplitudes (cross 
sections) of several semihard processes. It is shown that due to the presence 
of $\beta_0$-terms in the NLO expressions for the impact factors, the 
resulting optimal renormalization scale is not universal and depends both on 
the energy and on the type of process in question. We illustrate this 
general conclusion considering in details the following semihard processes:
\begin{itemize}
\item the forward amplitude of production of two light vector mesons in 
collision of two virtual photons, $\gamma^*\gamma^*\to V_1V_2$;
\item the high-energy behavior of the total cross section for highly virtual 
photons, $\gamma^*\gamma^*\to X$;
\item the inclusive production of two forward, 
high-$p_T$ jets separated by a large interval in rapidity $\Delta y$ 
(Mueller-Navelet jets), $p+p\to {\rm jet}+{\rm jet} +X$.
\end{itemize}

At present we do not have a model-independent method to resum the BFKL 
series beyond the NLA logarithms of the energy. Therefore we strictly adhere 
here to the original formulation of the BLM procedure and
do not consider its higher-order extensions, such as the 
\emph{sequential extended BLM}~\cite{Mikhailov:2004iq} and the
\emph{Principle of Maximum Conformality}~\cite{pmc} (see~\cite{Wu:2013ei} 
for a review on the latter method; see 
also~\cite{Wu:2014iba,Kataev:2014jba,Kataev:2014zwa,Ma:2015dxa} for some
recent comparisons between different optimization methods).

The paper is organized as follows: in section~2 we rederive the general
expression for the NLA BFKL amplitude in the $(\nu,n)$-representation;
in section~3 we discuss in detail the implementation of the BLM scale setting
method, both in exact way and in some approximated forms; in section~4 we 
present the applications of the procedure to the three different  
processes mentioned above; finally, in section~5 we draw our conclusions
and discuss previous studies of semihard processes with BLM method.

\section{The BFKL amplitude}

The cross section and many other physical observables are directly 
related to the forward amplitude, which in the BFKL approach can be 
expressed as follows:
\beq{ampl}
{\rm Im}_s
\left( {\cal A} \right)=\frac{s}{(2\pi)^{2}}\int\frac{d^{2}\vec q_1}{\vec
q_1^{\,\, 2}}\Phi_1(\vec q_1,s_0)\int
\frac{d^{2}\vec q_2}{\vec q_2^{\,\,2}} \Phi_2(-\vec q_2,s_0)
\int\limits^{\delta +i\infty}_{\delta
-i\infty}\frac{d\omega}{2\pi i}\left(\frac{s}{s_0}\right)^\omega
G_\omega (\vec q_1, \vec q_2)\, .
\eeq
This expression holds with NLA accuracy. Here, $s$ is the squared 
center-of-mass energy, whereas $s_0$ is an artificial scale introduced
to perform the Mellin transform from the $s$-space to the complex angular 
momentum plane and cancels in the full expression, up to terms beyond the NLA.
All momenta entering this expression are defined on the transverse plane
and are therefore two-dimensional.
$\Phi_{1,2}$ are the NLO impact factors specific of the process; we will see
later on three different examples for them. The Green's function $G_\omega$
takes care of the universal, energy-dependent part of the amplitude. It
obeys the BFKL equation
\begin{equation}
\omega \, G_\omega (\vec q_1,\vec q_2)  =\delta^{(2)} (\vec q_1-\vec q_2)
+\int d^{2}\vec q \, K(\vec q_1,\vec q) \,G_\omega (\vec q, \vec q_1) \;,
\end{equation}
where $K(\vec q_1,\vec q_2)$ is the BFKL kernel.

In this section we will derive a general form for the amplitude in the 
so-called $(\nu,n)$-representation, which will provide us with the
starting point of our further analysis. We will proceed along the same
lines of Refs.~\cite{mesons}. First of all, it is convenient to work in the 
transverse momentum representation, defined by
\beq{transv}
\hat{\vec q}\: |\vec q_i\rangle = \vec q_i|\vec q_i\rangle\;,
\eeq
\beq{norm}
\langle\vec q_1|\vec q_2\rangle =\delta^{(2)}(\vec q_1 - \vec q_2) \;,
\hspace{2cm}
\langle A|B\rangle =
\langle A|\vec k\rangle\langle\vec k|B\rangle =
\int d^2k A(\vec k)B(\vec k)\;.
\eeq
In this representation, the forward amplitude~(\ref{ampl}) takes the 
very simple form
\beq{ampl-transv}
{\rm Im}_s\left({\cal A}\right)=\frac{s}{(2\pi)^2}
\int_{\delta-i\infty}^{\delta+i\infty}\frac{d\omega}{2\pi i}
\, \left(\frac{s}{s_0}\right)^\omega
\langle\frac{\Phi_1}{\vec q_1^{\,\,2}}|\hat G_\omega|\frac{\Phi_2}{\vec q_2^{\,\,2}}
\rangle \ .
\eeq 

The kernel of the operator $\hat K$ becomes
\beq{kernel-op}
K(\vec q_2, \vec q_1) = \langle\vec q_2| \hat K |\vec q_1\rangle
\eeq
and the equation for the Green's function reads
\beq{Groper}
\hat 1=(\omega-\hat K)\hat G_\omega\;,
\eeq
its solution being
\beq{Groper1}
\hat G_\omega=(\omega-\hat K)^{-1} \, .
\eeq

The kernel is given as an expansion in the strong coupling,
\beq{kern}
\hat K=\bar \alpha_s \hat K^0 + \bar \alpha_s^2 \hat K^1\;,
\eeq
where
\beq{baral}
{\bar \alpha_s}=\frac{\alpha_s N_c}{\pi}
\eeq
and $N_c$ is the number of colors. In Eq.~(\ref{kern}) $\hat K^0$ is the
BFKL kernel in the LO, while $\hat K^1$ represents the NLO correction.

To determine the  cross section with NLA accuracy we need an
approximate solution of Eq.~(\ref{Groper1}). With the required accuracy this
solution is
\beq{exp}
\hat G_\omega=(\omega-\bar \alpha_s\hat K^0)^{-1}+
(\omega-\bar \alpha_s\hat K^0)^{-1}\left(\bar \alpha_s^2 \hat K^1\right)
(\omega-\bar \alpha_s \hat
K^0)^{-1}+ {\cal O}\left[\left(\bar \alpha_s^2 \hat K^1\right)^2\right]
\, .
\eeq

The basis of eigenfunctions of the LO kernel,
\beq{KLLA}
\hat K^0 |n,\nu\rangle = \chi(n,\nu)|n,\nu\rangle \, , \;\;\;\;\;\;\;\;\;\;
\chi (n,\nu)=2\psi(1)-\psi\left(\frac{n}{2}+\frac{1}{2}+i\nu\right)
-\psi\left(\frac{n}{2}+\frac{1}{2}-i\nu\right)\, ,
\eeq
is given by the following set of functions:
\beq{nuLLA}
\langle\vec q\, |n,\nu\rangle =\frac{1}{\pi\sqrt{2}}
\left(\vec q^{\,\, 2}\right)^{i\nu-\frac{1}{2}}e^{in\phi} \;,
\eeq
here $\phi$ is the azimuthal angle of the vector $\vec q$ counted from
some fixed direction in the transverse space, $\cos\phi \equiv q_x/|\vec q\,|$.
Then, the orthonormality and completeness conditions take the form
\beq{ort}
\langle n',\nu^\prime | n,\nu\rangle =\int \frac{d^2 \vec q}
{2 \pi^2 }\left(\vec q^{\,\, 2}\right)^{i\nu-i\nu^\prime -1}
e^{i(n-n')\phi}=\delta(\nu-\nu^\prime)\, \delta_{nn'}
\eeq
and
\beq{comp}
\hat 1 =\sum^{\infty}_{n=-\infty}\int\limits^{\infty}_{-\infty}d\nu \, 
| n,\nu\rangle\langle n,\nu |\ .
\eeq

The action of the full NLO BFKL kernel on these functions may be expressed
as follows:
\bea{Konnu}
\hat K|n,\nu\rangle &=&
\bar \alpha_s(\mu_R) \chi(n,\nu)|n,\nu\rangle
 +\bar \alpha_s^2(\mu_R)\left(\chi^{(1)}(n,\nu)
+\frac{\beta_0}{4N_c}\chi(n,\nu)\ln(\mu^2_R)\right)|n,\nu\rangle
\nonumber \\
&+& \bar
\alpha_s^2(\mu_R)\frac{\beta_0}{4N_c}\chi(n,\nu)
\left(i\frac{\partial}{\partial \nu}\right)|n,\nu\rangle \;,
\eea
where $\mu_R$ is the renormalization scale of the QCD coupling; the first
term represents the action of LO kernel, while the second and the third ones
stand for the diagonal and the non-diagonal parts of the NLO kernel and we
have used
\beq{beta00}
\beta_0=\frac{11 N_c}{3}-\frac{2 n_f}{3}\;,
\eeq
where $n_f$ is the number of active quark flavors.

The function $\chi^{(1)}(n,\nu)$, calculated in~\cite{Kotikov:2000pm} (see
also~\cite{Kotikov:2000pm2}), is conveniently represented in the form
\beq{ch11}
\chi^{(1)}(n,\nu)=-\frac{\beta_0}{8\, N_c}\left(\chi^2(n,\nu)-\frac{10}{3}
\chi(n,\nu)-i\chi^\prime(n,\nu)\right) + {\bar \chi}(n,\nu)\, ,
\eeq
where
\beq{chibar}
\bar \chi(n,\nu)\,=\,-\frac{1}{4}\left[\frac{\pi^2-4}{3}\chi(n,\nu)
-6\zeta(3)-\chi^{\prime\prime}(n,\nu) +\,2\,\phi(n,\nu)+\,2\,\phi(n,-\nu)
\right.
\eeq
\[
+ \left.
\frac{\pi^2\sinh(\pi\nu)}{2\,\nu\, \cosh^2(\pi\nu)}
\left(
\left(3+\left(1+\frac{n_f}{N_c^3}\right)\frac{11+12\nu^2}{16(1+\nu^2)}\right)
\delta_{n0}
-\left(1+\frac{n_f}{N_c^3}\right)\frac{1+4\nu^2}{32(1+\nu^2)}\delta_{n2}
\right)\right] \, ,
\]
\beq{phi}
\phi(n,\nu)\,=\,-\int\limits_0^1dx\,\frac{x^{-1/2+i\nu+n/2}}{1+x}
\left[\frac{1}{2}\left(\psi'\left(\frac{n+1}{2}\right)-\zeta(2)\right)
+\mbox{Li}_2(x)+\mbox{Li}_2(-x) \right.
\eeq
\[
\left. +\ln x \left(\psi(n+1)-\psi(1)+\ln(1+x)+\sum_{k=1}^\infty\frac{(-x)^k}
{k+n}\right)+\sum_{k=1}^\infty\frac{x^k}{(k+n)^2}(1-(-1)^k)\right]
\]
\[
=\sum_{k=0}^\infty\frac{(-1)^{k+1}}{k+(n+1)/2+i\nu}\left[\psi'(k+n+1)
-\psi'(k+1)+(-1)^{k+1}(\beta'(k+n+1)+\beta'(k+1))\right.
\]
\[
\left.
-\frac{1}{k+(n+1)/2+i\nu}(\psi(k+n+1)-\psi(k+1))\right] \, ,
\]
\[
\beta'(z)=\frac{1}{4}\left[\psi'\left(\frac{z+1}{2}\right)
-\psi'\left(\frac{z}{2}\right)\right]\;, \;\;\;\;\;
\mbox{Li}_2(x)=-\int\limits_0^xdt\,\frac{\ln(1-t)}{t} \, .
\]
Here and below $\chi^\prime(n,\nu)=d\chi(n,\nu)/d\nu$ and
$\chi^{\prime\prime}(n,\nu)=d^2\chi(n,\nu)/d^2\nu$.

The projection of the impact factors onto the eigenfunctions of the LO BFKL 
kernel, {\it i.e.} the transfer to the $(\nu,n)$-representation, is done as 
follows:
\[
\frac{\Phi_1(\vec q_1)}{\vec q_1^{\,\, 2}}=\sum^{+\infty}_{n=-\infty}
\int\limits^{+\infty}_{-\infty}
d\nu \, \Phi_1(\nu,n)\langle n,\nu| \vec q_1\rangle\, , \quad
\frac{\Phi_2(-\vec q_2)}{\vec q_2^{\,\, 2}}=\sum^{+\infty}_{n=-\infty}
\int\limits^{+\infty}_{-\infty} d\nu \, \Phi_2(\nu,n)
\langle \vec q_2 |n,\nu \rangle \, ,
\]
\beq{nu_rep}
\Phi_1(\nu,n)=\int d^2 q_1 \,\frac{\Phi_1(\vec q_1)}{\vec q_1^{\,\, 2}}
\frac{1}{\pi \sqrt{2}} \left(\vec q_1^{\,\, 2}\right)^{i\nu-\frac{1}{2}}
e^{i n \phi_1}\;,
\eeq
\[
\Phi_2(\nu,n)=\int d^2 q_2 \,\frac{\Phi_2(-\vec q_2)}{\vec q_2^{\,\, 2}}
\frac{1}{\pi \sqrt{2}} \left(\vec q_2^{\,\, 2}\right)^{-i\nu-\frac{1}{2}}
e^{-i n \phi_2}\;.
\]
The impact factors can be represented as an expansion in $\alpha_s$,
\beq{if}
\Phi_{1,2}(\vec q\,)=\alpha_s(\mu_R)\left[ v_{1,2}(\vec q\, )+ \bar \alpha_s(\mu_R)
v_{1,2}^{(1)}(\vec q\, )\right]
\eeq
and
\beq{vertex-exp}
\Phi_{1,2}(n,\nu)=\alpha_s(\mu_R)\left[ c_{1,2}(n,\nu)+ \bar \alpha_s(\mu_R)
c_{1,2}^{(1)}(n,\nu) \right]\, .
\eeq

To obtain our representation of the forward amplitude, we need the matrix 
element of the BFKL Green's function. According to~(\ref{exp}), we have
\[
\langle n,\nu|\hat G_\omega|n^\prime,\nu^\prime\rangle = \delta_{n,n^\prime}\left[
\delta(\nu-\nu^\prime)\left(
\frac{1}{\omega-\bar \alpha_s (\mu_R)\chi(n,\nu)}
\right.\right.
\]
\beq{Greens}
\left.
+\frac{\bar \alpha_s^2(\mu_R)(\bar \chi(n,\nu) 
+\frac{\beta_0}{8 N_c}(-\chi^2(n,\nu)+\frac{10}{3}\chi(n,\nu)+2\chi(n,\nu)
\ln \mu_R^2+i\frac{d}{d\nu}\chi(n,\nu)))}{(\omega-\bar \alpha_s(\mu_R) 
\chi(n,\nu))^2}\right)
\eeq
\[
\left.
+\frac{\frac{\beta_0}{4 N_c}\bar \alpha_s^2(\mu_R)\chi(n,\nu^\prime)}
{(\omega-\bar \alpha_s(\mu_R) \chi(n,\nu))(\omega-\bar \alpha_s(\mu_R) 
\chi(n,\nu^\prime))}\left(i\frac{d}{d\nu^\prime}\delta(\nu-\nu^\prime)\right)
\right]\ .
\]

Inserting twice the unity operator, written according to the completeness 
condition~(\ref{comp}), into~(\ref{ampl-transv}), we get
\[
{\rm Im}_s \left({\cal A}\right)=\frac{s}{(2\pi)^2}\sum^{\infty}_{n=-\infty}
\int\limits^{\infty}_{-\infty} d\nu\sum^{\infty}_{n^\prime =-\infty} 
\int\limits^{\infty}_{-\infty} d\nu^\prime \int_{\delta-i\infty}^{\delta+i\infty}
\frac{d\omega}{2\pi i}
\left(\frac{s}{s_0}\right)^\omega
\]
\beq{ampl-f}
\times\langle\frac{\Phi_1}{\vec q_1^{\,\,2}}|n,\nu\rangle\langle n,\nu|\hat 
G_\omega| n^\prime,\nu^\prime\rangle\langle n^\prime,\nu^\prime |
\frac{\Phi_2}{\vec q_2^{\,\,2}}\rangle \ ,
\eeq
and, after some algebra and integration by parts, finally
\[
{\rm Im}_s \left({\cal A}\right)=\frac{s}{(2\pi)^2}\sum^{\infty}_{n=-\infty}
\int\limits^{\infty}_{-\infty} d\nu \left(\frac{s}{s_0}\right)^{\bar \alpha_s(\mu_R)
\chi(n,\nu)} \alpha_s^2(\mu_R) c_1(n,\nu)c_2(n,\nu)
\]
\beq{ampl-ff}
\times\left[1+\bar \alpha_s(\mu_R)\left(\frac{c^{(1)}_1(n,\nu)}{c_1(n,\nu)}
+\frac{c^{(1)}_2(n,\nu)}{c_2(n,\nu)}\right) \right.
\eeq
\[
\left.
+\bar \alpha^2_s(\mu_R)\ln\frac{s}{s_0}\left\{\bar \chi(n,\nu)
+\frac{\beta_0}{8 N_c}\chi(n,\nu)\left(
-\chi(n,\nu)+\frac{10}{3}+2\ln \mu_R^2 +i\frac{d}{d\nu}\ln\frac{c_1(n,\nu)}
{c_2(n,\nu)}\right)\right\}\right] \, .
\]

This is our  {\it master}  representation of the NLA BFKL forward amplitude.
In the next section we will implement on it the BLM scale setting.  

\section{BLM scale setting}

The cross section of a process is related, via the optical theorem, to the 
imaginary part of the forward scattering amplitude,
\begin{equation}
\sigma =\frac{{\rm Im}_s A}{s} \ .
\end{equation}
Here we want to discuss the BLM scale setting for the separate contributions to 
the cross section, specified in~(\ref{ampl-ff}) by different values of $n$
and denoted in the following by ${\cal C}_n$.
Note that the $n=0$ case is relevant, {\it e.g.}, for the total cross sections 
of $\gamma^*\gamma^*$ interactions, Mueller-Navelet jet production and the 
forward differential cross section of the $\gamma^*\gamma^*\to V_1V_2$ process. 
Azimuthal angle correlations of produced jets in the Mueller-Navelet process 
are instead associated with non-zero values of $n$.

The starting point of our considerations is the expression for ${\cal C}_n$
in the $\overline{\rm MS}$ scheme (see Eq.~(\ref{ampl-ff})),
\[
{\cal C}_n
=\frac{1}{(2\pi)^2}\int\limits^{\infty}_{-\infty} d\nu 
\left(\frac{s}{s_0}\right)^{\bar \alpha_s(\mu_R)\chi(n,\nu)} \alpha_s^2(\mu_R) 
c_1(n,\nu)c_2(n,\nu)
\]
\beq{c_n}
\times\left[1+\bar \alpha_s(\mu_R)\left(\frac{c^{(1)}_1(n,\nu)}{c_1(n,\nu)}
+\frac{c^{(1)}_2(n,\nu)}{c_2(n,\nu)}\right) \right.
\eeq
\[
\left.
+\bar \alpha^2_s(\mu_R)\ln\frac{s}{s_0}\left\{\bar \chi(n,\nu) 
+\frac{\beta_0}{8 N_c}\chi(n,\nu)\left(
-\chi(n,\nu)+\frac{10}{3}+2\ln \mu_R^2 +i\frac{d}{d\nu}
\ln\frac{c_1(n,\nu)}{c_2(n,\nu)}\right)\right\}\right] \, .
\]
In the r.h.s. of this expression we have terms $\sim \alpha_s$ originated 
from the NLO corrections to the impact factors, and terms 
$\sim \alpha^2_s\ln(s/s_0)$ coming from NLO corrections to the BFKL kernel. In 
the latter case, the terms proportional to the QCD $\beta$-function are 
explicitly shown. For our further consideration of the BLM scale setting, 
similar contributions have to be separated also from the NLO impact factors.

In fact, the contribution to an NLO impact factor that is proportional to 
$\beta_0$ is universally expressed through the LO impact factor,
\beq{beta-if}
v^{(1)}(\vec q\,)=v(\vec q\,)\frac{\beta_0}{4 N_c}\left(\ln\left(\frac{\mu_R^2}
{\vec q\,^2}\right)+\frac{5}{3}\right)+\dots \ ,
\eeq
where the dots stand for the other terms, not proportional to $\beta_0$. This 
statement becomes evident if one considers the part of the strong coupling 
renormalization proportional to $n_f$ and related with the contributions of 
light quark flavors. Such contribution to the NLO impact factor originates only 
from diagrams with the light quark loop insertion in the Reggeized gluon 
propagator. The results for such contributions can be found, for instance, in   
Eq.~(5.1) of~\cite{Fadin:2001ap}. Tracing there the terms $\sim n_f$ and 
performing the QCD charge renormalization, one can indeed
confirm~(\ref{beta-if}).

Transforming~(\ref{beta-if}) to the $\nu$-representation according 
to~(\ref{nu_rep}), we obtain 
\bea{if2}
{\tilde{c}}_1^{\left(1\right)}(\nu, n)&=& \frac{\beta_0}{4 N_c}
\left[+i\frac{d}{d\nu} c_1(\nu,n)+\left(\ln \mu_R^2+\frac{5}{3}\right)
c_1(\nu, n)\right]\ ,
\nonumber
\\
{\tilde{c}}_2^{\left(1\right)}(\nu, n)&=&\frac{\beta_0}{4 N_c}
\left[-i\frac{d}{d\nu} c_2(\nu,n)+\left(\ln \mu_R^2+\frac{5}{3}\right)
c_2(\nu, n)\right] \ ,
\eea
and
\beq{}
\frac{{\tilde{c}}_1^{\left(1\right)}}{c_1}+\frac{{\tilde{c}}_2^{\left(1\right)}}{c_2}
=\frac{\beta_0}{4 N_c}\left[i\frac{d}{d\nu}\ln\left(\frac{c_1}{c_2}\right)
+2\left(\ln \mu_R^2+\frac{5}{3}\right)\right] \ .
\eeq

It is convenient to introduce the function $f\left(\nu\right)$, defined
through
\beq{}
i\frac{d}{d\nu}\ln\left(\frac{c_1}{c_2}\right)\equiv 2 \left[f(\nu)
-\ln\left(Q_1 Q_2\right)\right]\ ,
\eeq
that depends on the given process, where $Q_{1,2}$ denote here the hard scales 
which enter the impact factors $c_{1,2}$~\footnote{Here we consider processes 
whose impact factors are characterized by only one hard scale. This is the 
virtuality of the photon, $Q$, for the $\gamma^*\to\gamma^*$ and 
$\gamma^*\to V$ impact factors, and the jet transverse momentum, $|\vec k|$,
for the impact factor describing the Mueller-Navelet jet production.}. 
The specific form of the function $f(\nu)$ depends on the particular process.
According to the properties of the corresponding LO impact factors 
($\gamma^*\to V$, $\gamma^*\to\gamma^*$ and Mueller-Navelet jet vertex), one 
can easily check that
\begin{equation}
f_{\gamma^*\gamma^*\to X}(\nu)=f_{pp\to {\rm jet}_1+X+{\rm jet}_2}(\nu)=0 \ ,
\label{fgammajet}
\end{equation} 
for the processes $\gamma^*\gamma^*\to X$ and Mueller-Navelet jet production,
whereas for the process $\gamma^*\gamma^*\to V_1V_2$ (forward electroproduction 
of two light vector mesons) this function is not equal to zero,
\beq{fVV}
f_{\gamma^*\gamma^*\to V_1V_2}\left(\nu\right)=\psi\left(3+2 i\nu\right) 
+ \psi\left(3-2 i\nu\right) - \psi\left(\frac{3}{2}+ i\nu\right) 
- \psi\left(\frac{3}{2}- i\nu\right)\, .
\eeq

Now, we present again our result for the generic observable ${\cal C}_n$, 
showing explicitly all contributions proportional to the QCD $\beta$-function,
{\it i.e.} also those originating from the impact factors:
\[
{\cal C}_n
=\frac{1}{(2\pi)^2}\int\limits^{\infty}_{-\infty} d\nu \left(\frac{s}{s_0}
\right)^{\bar \alpha_s(\mu_R)\chi(n,\nu)} \alpha_s^2(\mu_R) c_1(n,\nu)c_2(n,\nu)
\]
\beq{c_nn}
\times\left[1+\bar \alpha_s(\mu_R)\left(\frac{\bar c^{(1)}_1(n,\nu)}{c_1(n,\nu)}
+\frac{\bar c^{(1)}_2(n,\nu)}{c_2(n,\nu)}
+\frac{\beta_0}{2 N_c}\left(\frac{5}{3}+\ln \frac{\mu_R^2}{Q_1 Q_2} +f(\nu)
\right)\right)\right.
\eeq
\[
\left.
+\bar \alpha^2_s(\mu_R)\ln\frac{s}{s_0}\left\{\bar \chi(n,\nu)
+\frac{\beta_0}{4 N_c}\chi(n,\nu)\left(
-\frac{\chi(n,\nu)}{2}+\frac{5}{3}+\ln \frac{\mu_R^2}{Q_1 Q_2} +f(\nu)\right)
\right\}\right] \, ,
\]
where $\bar c^{(1)}_{1,2} \equiv c^{(1)}_{1,2}- \tilde c^{(1)}_{1,2}$. We note that 
the dependence of~(\ref{c_nn}) on the scale $\mu_R$ is subleading:
performing in~(\ref{c_nn}) the replacement
\beq{alphaSrun}
\alpha_s(\mu_R)=\alpha_s(\mu^\prime_R)\left(1-\bar\alpha_s(\mu^\prime_R)
\frac{\beta_0}{2N_c}\ln\frac{\mu_R}{\mu^\prime_R}\right) \, ,
\eeq
one indeed obtains the same expression as before with the
new scale $\mu_R^\prime$ at the place of the old one $\mu_R$, 
plus some additional contributions which are beyond the NLA accuracy.

As the next step, we perform a finite renormalization from the 
$\overline{\rm MS}$ to the physical MOM scheme, that means:
\beq{scheme}
\alpha_s^{\overline{\rm MS}}=\alpha_s^{\rm MOM}\left(1+\frac{\alpha_s^{\rm MOM}}{\pi}T 
\right)\;,
\eeq
with $T=T^{\beta}+T^{\rm conf}$,
\beq{}
T^{\beta}=-\frac{\beta_0}{2}\left( 1+\frac{2}{3}I \right)\, ,
\eeq
\[
T^{\rm conf}= \frac{C_A}{8}\left[ \frac{17}{2}I +\frac{3}{2}\left(I-1\right)\xi
+\left( 1-\frac{1}{3}I\right)\xi^2-\frac{1}{6}\xi^3 \right] \;,
\]
where $I=-2\int_0^1dx\frac{\ln\left(x\right)}{x^2-x+1}\simeq2.3439$ and $\xi$ 
is a gauge parameter, fixed at zero in the following.

Inserting~(\ref{scheme}) into~(\ref{c_nn}) and expanding the result, 
we obtain, within NLA accuracy,
\[
{\cal C}^{\rm MOM}_n =\frac{1}{(2\pi)^2}\int\limits^{\infty}_{-\infty} 
d\nu \left(\frac{s}{s_0}\right)^{\bar \alpha^{\rm MOM}_s(\mu_R)\chi(n,\nu)} 
\left(\alpha^{\rm MOM}_s (\mu_R)\right)^2 c_1(n,\nu)c_2(n,\nu)
\]
\[
\times\left[1+\bar \alpha^{\rm MOM}_s(\mu_R)\left\{\frac{\bar c^{(1)}_1(n,\nu)}
{c_1(n,\nu)}+\frac{\bar c^{(1)}_2(n,\nu)}{c_2(n,\nu)}+\frac{2T^{\rm conf}}{N_c}
\right.\right.
\]
\[
\left.
+\frac{\beta_0}{2 N_c}\left(\frac{5}{3}+\ln \frac{\mu_R^2}{Q_1 Q_2} +f(\nu)
-2\left( 1+\frac{2}{3}I \right)\right)
\right\}
\]
\beq{c_nnn}
+\left(\bar \alpha^{\rm MOM}_s(\mu_R)\right)^2\ln\frac{s}{s_0}
\left\{\bar \chi(n,\nu) +\frac{T^{\rm conf}}{N_c}\chi(n,\nu)\right.
\eeq
\[
\left.\left.
+\frac{\beta_0}{4 N_c}\chi(n,\nu)\left(
-\frac{\chi(n,\nu)}{2}+\frac{5}{3}+\ln \frac{\mu_R^2}{Q_1 Q_2} +f(\nu)
-2\left(1+\frac{2}{3}I \right)\right)\right\}\right] \, .
\]
The optimal scale $\mu_R^{\rm BLM}$ is the value of $\mu_R$ that makes the 
expression proportional to $\beta_0$ vanish. We thus have
\[
{\cal C}^{\beta}_n
=\frac{1}{(2\pi)^2}\int\limits^{\infty}_{-\infty} d\nu 
\left(\frac{s}{s_0}\right)^{\bar \alpha^{\rm MOM}_s(\mu^{\rm BLM}_R)\chi(n,\nu)} 
\left(\alpha^{\rm MOM}_s (\mu^{\rm BLM}_R)\right)^3
\]
\beq{c_nnnbeta}
\times c_1(n,\nu)c_2(n,\nu) \frac{\beta_0}{2 N_c} \left[\frac{5}{3}
+\ln \frac{(\mu^{\rm BLM}_R)^2}{Q_1 Q_2} +f(\nu)-2\left( 1+\frac{2}{3}I \right)
\right.
\eeq
\[
\left.
+\bar \alpha^{\rm MOM}_s(\mu^{\rm BLM}_R)\ln\frac{s}{s_0} \: \frac{\chi(n,\nu)}{2}
\left(-\frac{\chi(n,\nu)}{2}+\frac{5}{3}+\ln \frac{(\mu^{\rm BLM}_R)^2}{Q_1 Q_2} 
+f(\nu)-2\left( 1+\frac{2}{3}I \right)\right)\right]=0 \, .
\]
In the r.h.s. of~(\ref{c_nnnbeta}) we have two groups of contributions. The 
first one originates from the $\beta_0$-dependent part of NLO impact 
factor~(\ref{beta-if}) and also from the expansion of the common 
$\alpha^2_s$ pre-factor in~(\ref{c_nn}) after expressing it in terms of 
$\alpha_s^{\rm MOM}$.
The other group are the terms proportional to $\bar \alpha_s^{\rm MOM}\ln s/s_0$. 
These contributions are those $\beta_0$-dependent terms that are proportional 
to $\ln s/s_0$ in~(\ref{c_nn}) and also the one coming from the expansion of  
the $(s/s_0)^{\bar \alpha_s \chi(n,\nu)}$ factor in~(\ref{c_nn}) after expressing it 
in terms of $\alpha_s^{\rm MOM}$.

The solution of Eq.~(\ref{c_nnnbeta}) gives us the value of BLM scale. Note 
that this solution depends on the energy (on the ratio $s/s_0$).
Such scale setting procedure is a direct application of the original BLM 
approach to semihard processes. Finally, our expression for the observable 
reads
\beq{c_BLMmain}
{\cal C}^{\rm BLM}_n
=\frac{1}{(2\pi)^2}\int\limits^{\infty}_{-\infty} d\nu \left(\frac{s}{s_0}
\right)^{\bar \alpha^{\rm MOM}_s(\mu^{\rm BLM}_R)\left[\chi(n,\nu)
+\bar \alpha^{\rm MOM}_s(\mu^{\rm BLM}_R)\left(\bar \chi(n,\nu) +\frac{T^{\rm conf}}
{N_c}\chi(n,\nu)\right)\right]}
\eeq
\[
\times \left(\alpha^{\rm MOM}_s (\mu^{\rm BLM}_R)\right)^2 c_1(n,\nu)c_2(n,\nu) 
\left[1+\bar \alpha^{\rm MOM}_s(\mu^{\rm BLM}_R)\left\{\frac{\bar c^{(1)}_1(n,\nu)}
{c_1(n,\nu)}+\frac{\bar c^{(1)}_2(n,\nu)}{c_2(n,\nu)}+\frac{2T^{\rm conf}}{N_c}
\right\} \right] \, ,
\]
where we put at the exponent the terms $\sim \bar \alpha_s^{\rm MOM}\ln s/s_0$,
which is allowed within the NLA accuracy.

Unfortunately, Eq.~(\ref{c_nnnbeta}) can be solved only numerically, thus
making the scale setting a bit unpractical. For this reason, we will work
out also some analytic approximate approaches to the BLM scale setting, which
have the merit of a straightforward and simple application.
We consider the BLM scale as a function of $\nu$ and chose it in order to make
vanish either the first or the second ($\sim \bar \alpha_s^{\rm MOM}\ln s/s_0$) 
group of terms in the Eq.~(\ref{c_nnnbeta}). We thus have two cases:
\begin{itemize}
\item case $(a)$
\beq{casea}
\left(\mu_{R, a}^{\rm BLM}\right)^2=Q_1Q_2\ \exp\left[2\left(1+\frac{2}{3}I\right)
-f\left(\nu\right)-\frac{5}{3}\right]\ ,
\eeq
\[
{\cal C}^{\rm BLM, a}_n
=\frac{1}{(2\pi)^2}\int\limits^{\infty}_{-\infty} d\nu \left(\frac{s}{s_0}
\right)^{\bar \alpha^{\rm MOM}_s(\mu^{\rm BLM}_{R, a})\left[\chi(n,\nu)+\bar \alpha^{\rm MOM}_s
(\mu^{\rm BLM}_{R, a})\left(\bar \chi(n,\nu) +\frac{T^{\rm conf}}{N_c}\chi(n,\nu)
-\frac{\beta_0}{8 N_c}\chi^2(n,\nu)\right)\right]}
\]
\beq{c_BLMa}
\times \left(\alpha^{\rm MOM}_s (\mu^{\rm BLM}_{R, a})\right)^2 c_1(n,\nu)c_2(n,\nu) 
\eeq
\[
\times \left[1+\bar \alpha^{\rm MOM}_s(\mu^{\rm BLM}_{R, a})
\left\{\frac{\bar c^{(1)}_1(n,\nu)}{c_1(n,\nu)}+\frac{\bar c^{(1)}_2(n,\nu)}
{c_2(n,\nu)}+\frac{2T^{\rm conf}}{N_c}
\right\} \right] \, ,
\]
\item case $(b)$
\beq{caseb}
\left(\mu_{R, b}^{\rm BLM}\right)^2=Q_1Q_2\ \exp\left[2\left(1+\frac{2}{3}I\right)
-f\left(\nu\right)-\frac{5}{3}+\frac{1}{2}\chi\left(\nu,n\right)\right]\ ,
\eeq
\[
{\cal C}^{\rm BLM, b}_n
=\frac{1}{(2\pi)^2}\int\limits^{\infty}_{-\infty} d\nu \left(\frac{s}{s_0}
\right)^{\bar \alpha^{\rm MOM}_s(\mu^{\rm BLM}_{R, b})\left[\chi(n,\nu)+\bar \alpha^{\rm MOM}_s
(\mu^{\rm BLM}_{R, b})\left(\bar \chi(n,\nu) +\frac{T^{\rm conf}}{N_c}\chi(n,\nu)
\right)\right]}
\]
\beq{c_BLMb}
\times \left(\alpha^{\rm MOM}_s (\mu^{\rm BLM}_{R, b})\right)^2 c_1(n,\nu)c_2(n,\nu)
\eeq
\[
\times\left[1+\bar \alpha^{\rm MOM}_s(\mu^{\rm BLM}_{R, b})\left\{\frac{\bar c^{(1)}_1
(n,\nu)}{c_1(n,\nu)}+\frac{\bar c^{(1)}_2(n,\nu)}{c_2(n,\nu)}
+\frac{2T^{\rm conf}}{N_c}+\frac{\beta_0}{4 N_c}\chi(n,\nu)
\right\}\right]\, .
\]
\end{itemize}

The other possible option for the BLM scale setting could be related with the 
requirement that the entire expression in the integrand of~(\ref{c_nnnbeta})
vanishes, which leads to the following condition
\begin{itemize}
\item case $(c)$
\beq{casec}
\frac{5}{3}+\ln \frac{(\mu^{\rm BLM}_{R, c})^2}{Q_1 Q_2} +f(\nu)
-2\left( 1+\frac{2}{3}I \right)= 
\frac{\bar \alpha^{\rm MOM}_s(\mu^{\rm BLM}_{R, c})\ln\frac{s}{s_0} \:
\frac{\chi^2(n,\nu)}{4}}{1+\bar \alpha^{\rm MOM}_s(\mu^{\rm BLM}_{R, c})
\ln\frac{s}{s_0} \: \frac{\chi(n,\nu)}{2}} \, .
\eeq
\end{itemize}
One should mention, however, that such approach to the BLM scale setting has 
a limited applicability, since the denominator in the r.h.s. of~(\ref{casec}) 
vanishes at some value of $\nu=\bar \nu$, given by 
\beq{barnu}
1+\bar \alpha^{MOM}_s\ln\frac{s}{s_0} \frac{\chi(n,\bar \nu)}{2}=0 \, ,
\eeq
which prevents us from defining $\mu^{\rm BLM}_{R, c}(\nu)$ in the entire 
$\nu$ range. Nevertheless, one can try to use such method in those cases
when the product of the two LO impact factors $c_1(n,\nu) c_2(n,\nu)$ is a 
function decreasing so rapidly to guarantee the convergence of the 
$\nu$-integration in~(\ref{c_BLMmain}) in the $\nu$-region where there is no 
problem with the solution of Eq.~(\ref{casec}).

Note also that all three approaches to BLM scale fixing discussed above, and 
given in Eqs.~(\ref{casea}), (\ref{caseb}) and~(\ref{casec}), could be 
applicable only to processes characterized by a real-valued function 
$f(\nu)$. For some processes this is not the case. In particular, the inclusive 
production of two identified hadrons separated by large interval of rapidity in 
proton-proton collisions, $p+p\to h_1+h_2 +X$, is described by a complex-valued 
function, $f^*(\nu)=f(-\nu)$. This can be easily seen calculating $f(\nu)$ 
from Eq.~(77) of~\cite{Ivanov:2012iv} for the identified hadron production 
impact factor. In such cases one can use only the BLM scale fixing method 
which relies on the numerical solution of Eq.~(\ref{c_nnnbeta}).    

\section{Applications}

In this section we apply the BLM approach to a selection of semihard processes. 
For the energy variables we will use notations
\beq{YY0}
Y=\ln\frac{s}{Q^2} \, , \quad\quad\quad Y_0=\ln\frac{s_0}{Q^2} \, . 
\eeq   
In our numerics we use the following settings: $n_f=5$ and 
$\alpha_s(M_Z)=0.11707$ for the number of active flavors and the value of the 
strong coupling.

\subsection{Electroproduction of two vector mesons}

We start with the description of the forward amplitude for the production of a 
pair of light vector mesons in the collision of two virtual photons, 
$\gamma^*\gamma^*\to V_1V_2$. Such processes could be studied in experiments at 
future high-energy $e^+e^-$ colliders, 
see~\cite{Pire:2005ic,Segond:2007fj,Goncalves:2006wy} for estimates of the 
cross section in the Born approximation. The BFKL resummation for these 
processes was considered in~\cite{Enberg:2005eq}, where the inclusion of 
NLO effects was limited to the corrections to the BFKL kernel. In the
papers~\cite{mesons}, some of us performed a complete NLA BFKL analysis for 
the forward amplitude of these processes, including the NLO corrections also 
to the $\gamma^*\to V$ impact factors~\cite{Ivanov:2004pp}. Very large NLA 
corrections to the forward amplitude were found, therefore in~\cite{mesons} 
the \emph{Principle of Minimal Sensitivity} (PMS)~\cite{PMS}
approach was used to optimize the perturbative series. 

\begin{figure}[t]
\centering
\begin{minipage}{0.50\textwidth}
\phantom{.}\vspace{0.2cm}
\includegraphics[scale=0.47]{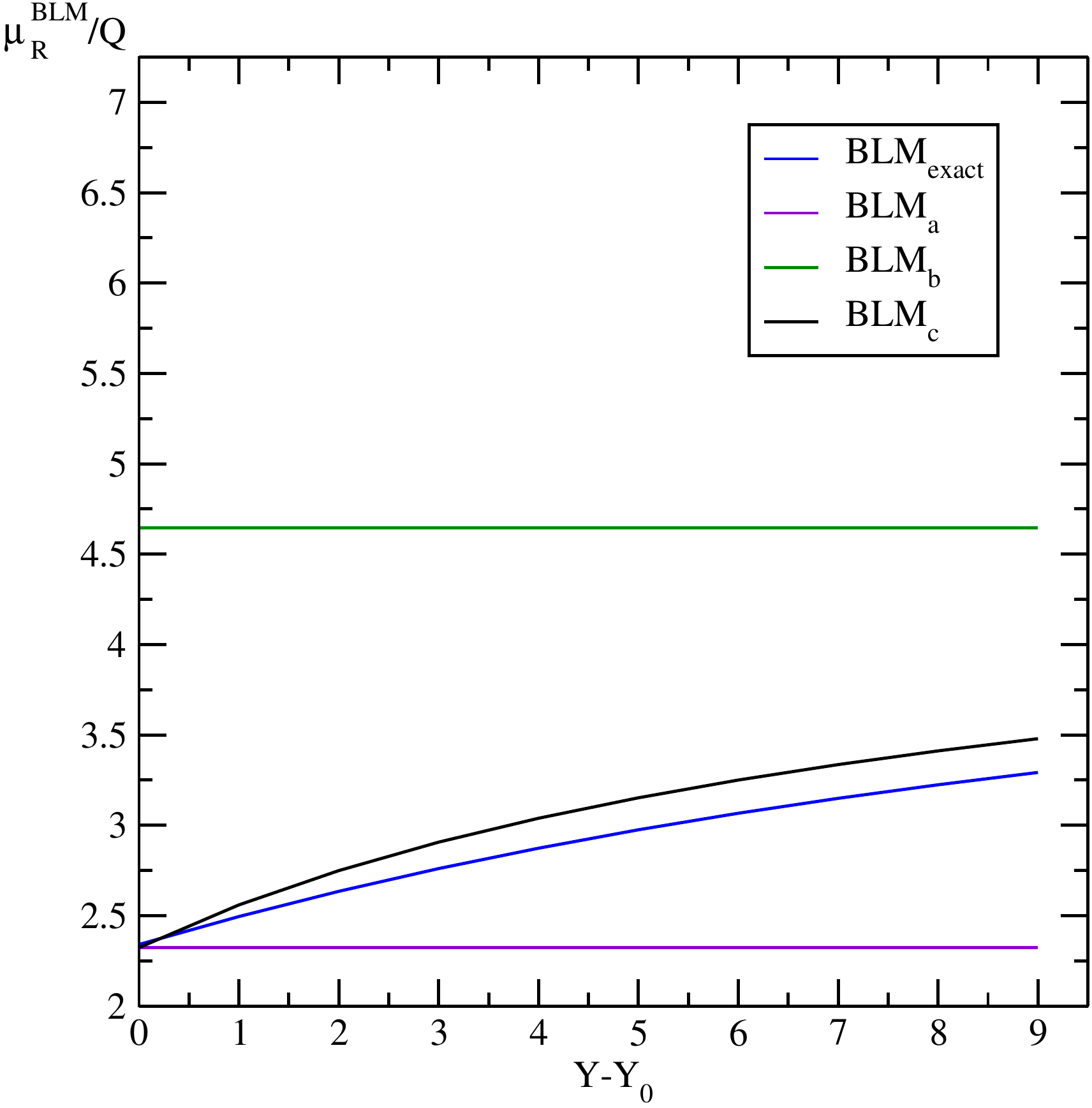}\hspace{0.3cm}
\end{minipage}
\begin{minipage}{0.48\textwidth}
\includegraphics[scale=0.47]{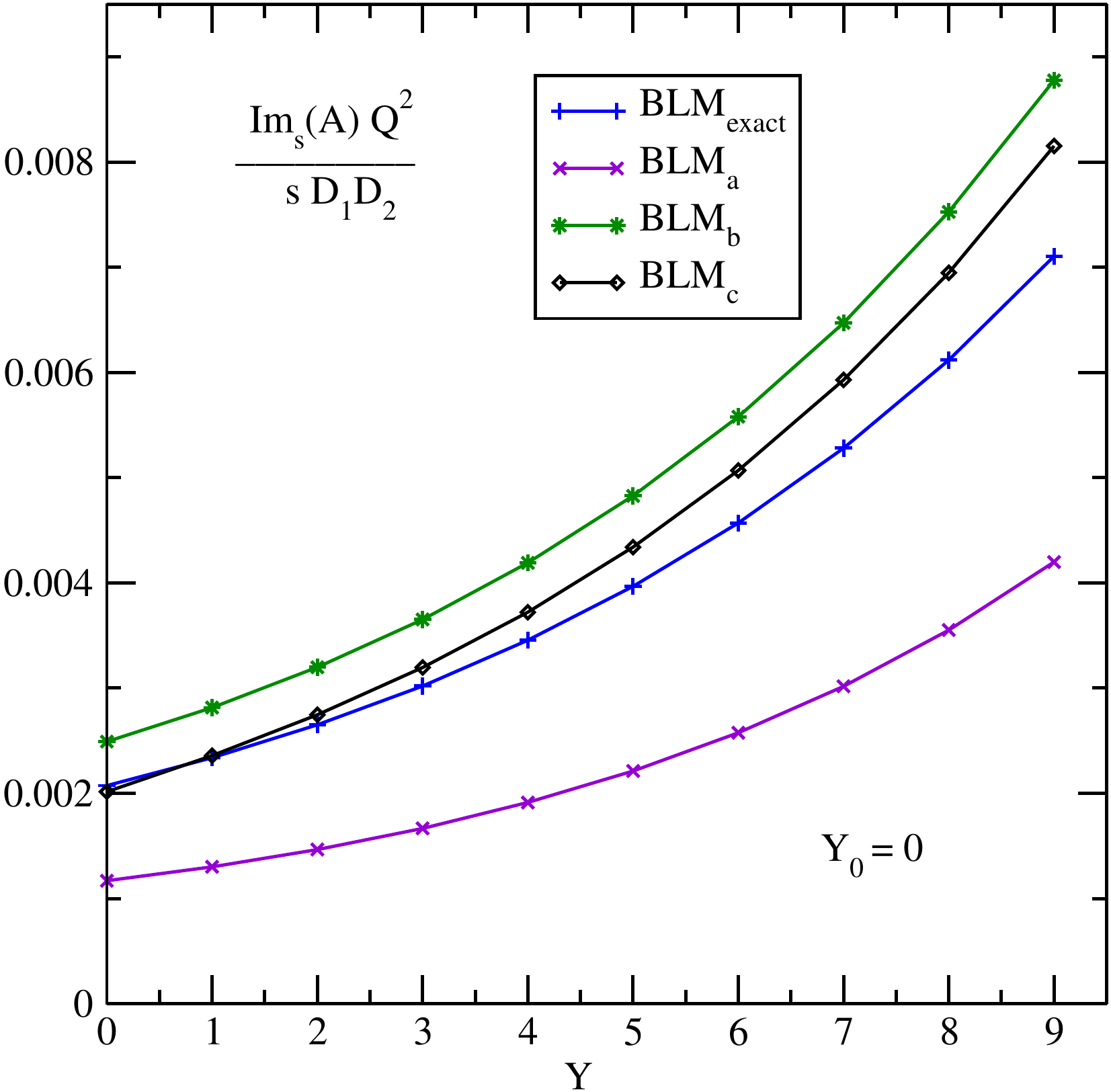}
\end{minipage}
\vspace{-0.2cm}
\caption[]{Left: BLM scales for the process $\gamma^*\gamma^*\to V_1V_2$ 
(see the text for details). Right: Forward amplitude for $\gamma^*\gamma^*\to 
V_1V_2$ at $Y_0=0$.}
\label{fig:mesons}
\end{figure}

Here we present numerical results for the forward amplitude obtained with the
BLM optimization method described above.      

We consider the case of equal values of photons virtualities, $Q_1=Q_2=Q$, and,
following the first of Refs.~\cite{mesons}, present our numerical predictions 
for the forward amplitude multiplied by some kinematic factors,  ${\rm Im}_s
\left({\cal A}\right)Q^2/(s D_1 D_2)$, calculated at $Q=50$~GeV, where the
expressions for $D_{1,2}$ are given in Eq.~(14) of the first of 
Refs.~\cite{mesons}.

For the considered process only the $n=0$ term contributes and the 
$f(\nu)$ function is given in~(\ref{fVV}). We will try all approaches to the 
BLM scale setting described in previous section. In particular, for this process
the product of two LO impact factors $c_1(n,\nu)c_2(n,\nu)$ vanishes very fast 
for $|\nu|>1$, therefore in the relevant integration $\nu$-range, 
$|\nu|<1$, we can find the solution of Eq.~(\ref{casec}) and determine the
BLM scale $\mu_{R,c}^{\rm BLM}$ as a function of $\nu$ and energy.  

In Fig.~\ref{fig:mesons}(left) we show the values of the BLM to kinematic 
scale ratios, $\mu_R^{\rm BLM}/Q$, as functions of $Y-Y_0$, obtained in four 
different cases. By ``exact'' case we denote the scale obtained solving 
numerically Eq.~(\ref{c_nnnbeta}) for each value of $\ln\left(s/s_0\right)\equiv
Y-Y_0$. In the other three approaches, the BLM scales depend on 
$\nu$: the scales for cases~$(a)$ and $(b)$ are given by Eqs.~(\ref{casea}) 
and~(\ref{caseb}), respectively; the case~$(c)$ corresponds to the numerical 
solution of Eq.~(\ref{casec}) for each values of $\nu$ and $Y-Y_0$.  The 
$\nu$-dependent scales, cases~$(a)$, $(b)$ and~$(c)$ are shown in 
Fig.~\ref{fig:mesons}(left) for the particular value of $\nu=0$.

Approximate approaches to the scale setting give energy-independent BLM scales
(see cases~$(a)$ and $(b)$ in Fig.~\ref{fig:mesons}(left)), whereas an exact 
implementation of the BLM rule leads in general to the scales which depend on 
the energy of the process (see cases~$(c)$ and ``exact'' in 
Fig.~\ref{fig:mesons}(left)). In fact, the approaches $(a)$ and $(b)$ can be 
considered as a low- and a high-energy approximation to the case~$(c)$, where 
the BLM scale setting prescription is implemented precisely. 

Nevertheless, as we already mentioned above, the condition~(\ref{casec}) could 
not be resolved for all processes. Therefore we defined also a method which 
could be universally applied and which we call here ``exact''. It gives a 
$\nu$-independent BLM scale and it is based on the requirement of vanishing 
of the {\em integral} in Eq.~(\ref{c_nnnbeta}), contrary to the approach $(c)$, 
where we try to make vanish the {\em integrand} of the same equation for each 
separate value of $\nu$. 

In Fig.~\ref{fig:mesons}(right) we show our predictions as functions of the
energy for the forward amplitude calculated with all the four different methods 
described above: cases~$(a)$ and $(b)$ were calculated using 
Eqs.~(\ref{c_BLMa}) and~(\ref{c_BLMb}), cases~$(c)$ and ``exact'' 
using Eq.~(\ref{c_BLMmain}) with the corresponding choices of the scales. The 
result of the BFKL resummation depends not only on the renormalization scale 
$\mu_R$, which is fixed here with the BLM method, but also on the energy scale 
$s_0$ or $Y_0$. In Fig.~\ref{fig:mesons}(right) we present the results obtained 
with the choice of this scale dictated by the kinematic of the process, 
$s_0=Q^2$ or $Y_0=0$.  
A more reliable estimation could result from fixing the value of 
$Y_0$ according to some optimization method, such as PMS, but this goes 
beyond the scope of present paper.

As we can see in Fig.~\ref{fig:mesons}(right), our predictions obtained with 
precise implementations of BLM method lie inbetween those derived with the use 
of the two approximate realizations. Note that the difference between the two 
explicit methods, cases~$(c)$ and ``exact'', is sizeable and increases with the 
energy. 
This is related to the fact that these two approaches are not equivalent,
and the scales in the case~$(c)$ are larger than those in the ``exact'' one.  
Note also that, with the growth of energy, the value of $\nu=\bar \nu$ where 
the solution of Eq.~(\ref{casec}) has a singularity decreases, see 
Eq.~(\ref{barnu}), and approaches the $\nu$-range important for the 
determination of our observable.      

\begin{figure}[t]
\centering
\includegraphics[scale=0.47]{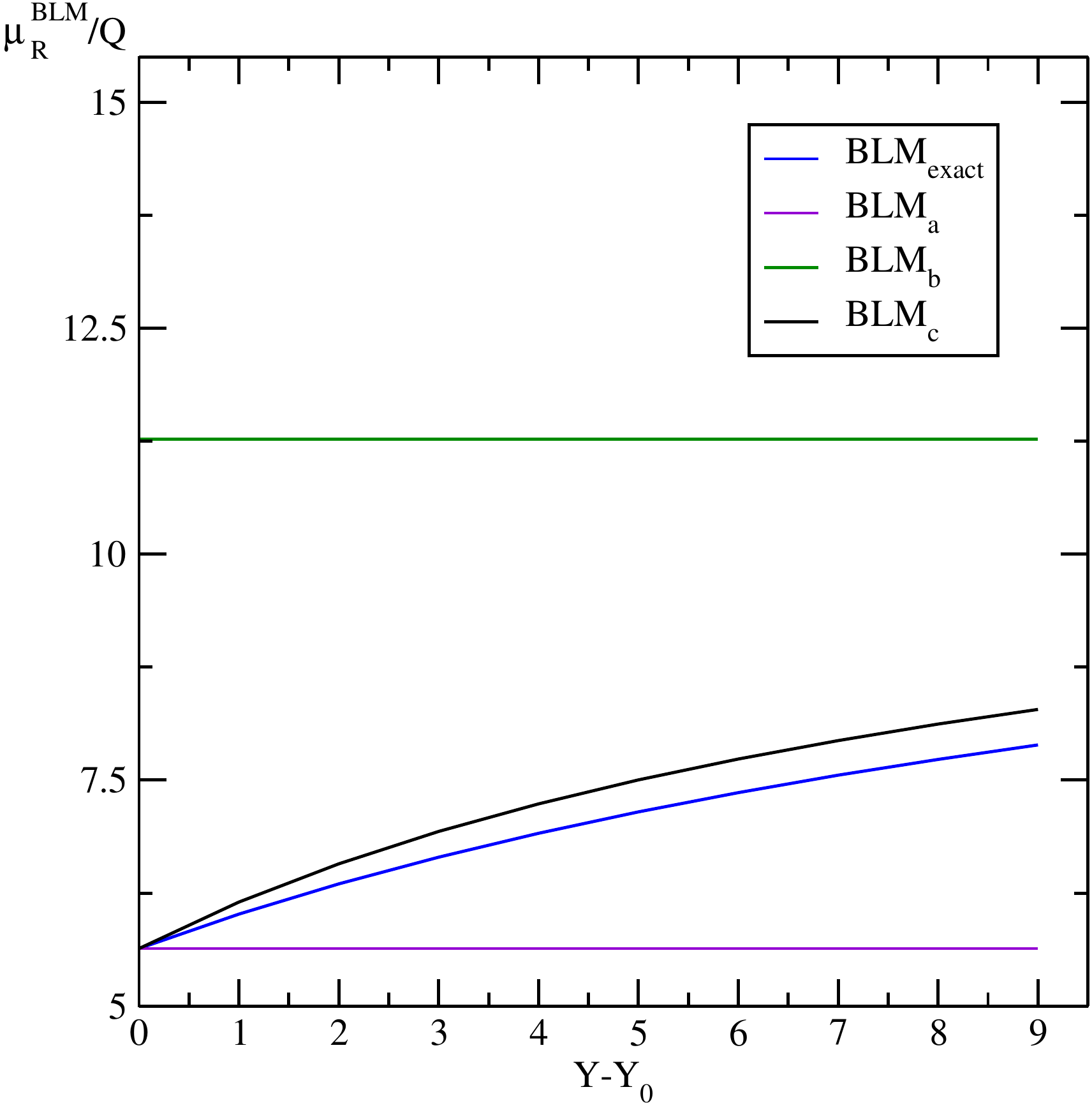}
\caption[]{BLM scales for the process $\gamma^*\gamma^*\to X$ 
(see the text for details).}
\label{fig:photons}
\end{figure}

\subsection{$\gamma^*\gamma^*$ total cross section}

In~\cite{Ivanov:2014hpa} some of us studied the $\gamma^*\gamma^*$ total cross 
section in the NLA BFKL approach considering two different optimization methods 
of the perturbative series. One of them was the BLM method, cases~$(a)$ and 
$(b)$, described above, where the Eqs.~(\ref{c_BLMa}) and~(\ref{c_BLMb}) were 
transformed back to $\overline{\rm MS}$ scheme.  In that paper we fixed
the photon virtualities and, correspondingly, the number of active flavors 
$n_f$ in order to make a comparison with LEP2 experimental data.
Here, we are interested in the general features of the BLM scale setting
procedure, therefore we prefer to fix the photon virtualities as in the
two-meson production: $Q_1=Q_2\equiv Q=50$ GeV with $n_f=5$. 

In Fig.~\ref{fig:photons}, as in the case of the vector mesons, we show the 
four different ratios $\mu_R^{\rm BLM}/Q$ versus $Y-Y_0$. The four cases~$(a)$, 
$(b)$, $(c)$ and ``exact'' are defined exactly as in the previous subsection.
As we mentioned before, cases~$(a)$ and~$(b)$ are independent on the energy
of the process, but depend on the kind of process through the $f(\nu)$ 
function. In particular, for the production of a pair of light vector mesons 
the function is given by Eq.~(\ref{fVV}), while for this process it is 
$f(\nu)=0$ (see Eq.~(\ref{fgammajet})).  

For this process we only discuss here the BLM scale setting and 
do not present its cross section. The $\gamma^*\gamma^*$ cross section was 
already considered in~\cite{Ivanov:2014hpa}, where serious problems were 
found, related with the very large values of NLO 
corrections~\cite{Balitsky:2012bs} for the virtual photon impact factor. 
For details and an extended discussion of this issue, we refer the reader 
to~\cite{Ivanov:2014hpa}.  

\subsection{Mueller-Navelet jets}

The last semihard process that we consider is the production of two forward 
high-$p_T$ jets produced with a large separation in rapidity $\Delta y$ 
(Mueller-Navelet jets~\cite{Mueller:1986ey}). Such process was studied at 
Large Hadron Collider (LHC): the CMS collaboration provided with data for 
azimuthal decorrelations~\cite{CMS} that can be expressed, from a theoretical 
point of view, by ratios ${\cal {C}}_m/{\cal{C}}_n$, where ${\cal{C}}_n$ are 
to be averaged over $k_{J_i}$ (jet transverse momentum) and $y_{J_i}$ 
(rapidity jet). Then, in order to match the kinematic cuts used by the CMS 
collaboration, we have
\begin{eqnarray}
C_n=\int_{y_{1,\rm min}}^{y_{1,\rm max}}dy_1
\int_{y_{2,\rm min}}^{y_{2,\rm max}}dy_2\int_{k_{J_1,\rm min}}^{\infty}dk_{J_1}
\int_{k_{J_2,\rm min}}^{\infty}dk_{J_2} \delta\left(y_1-y_2-Y\right){\cal C}_n
\left(y_{J_1},y_{J_2},k_{J_1},k_{J_2} \right)\;,
\end{eqnarray}
with $y_{1,\rm min}=y_{2,\rm min}=-4.7$, 
$y_{1,\rm max}=y_{2,\rm max}=4.7$~\footnote{In~\cite{Caporale2014} it was 
mistakenly written $y_{i,\rm min}=0$, although all numerical results presented
there were obtained using the correct value $y_{i,\rm min}=-4.7$.} and 
$k_{J_1,\rm min}=k_{J_2,\rm min}=35$ GeV. The comparison between experimental results
for jets with cone radius $R=0.5$ produced at a center-of-mass energy of 
$\sqrt s=7$ TeV and theoretical calculations was done 
in~\cite{Ducloue2014}, where the exact NLO impact factors calculated 
in~\cite{IFjet} were used, and in~\cite{Caporale2014}, where the NLO impact
factors were taken in the small-cone approximation as calculated
in~\cite{SCA}~\footnote{ For a critical comparison of the different
expressions for the forward jet vertex, we refer to~\cite{Colferai:2015zfa}.
}. 

\begin{figure}[t]
\centering
\includegraphics[scale=0.45]{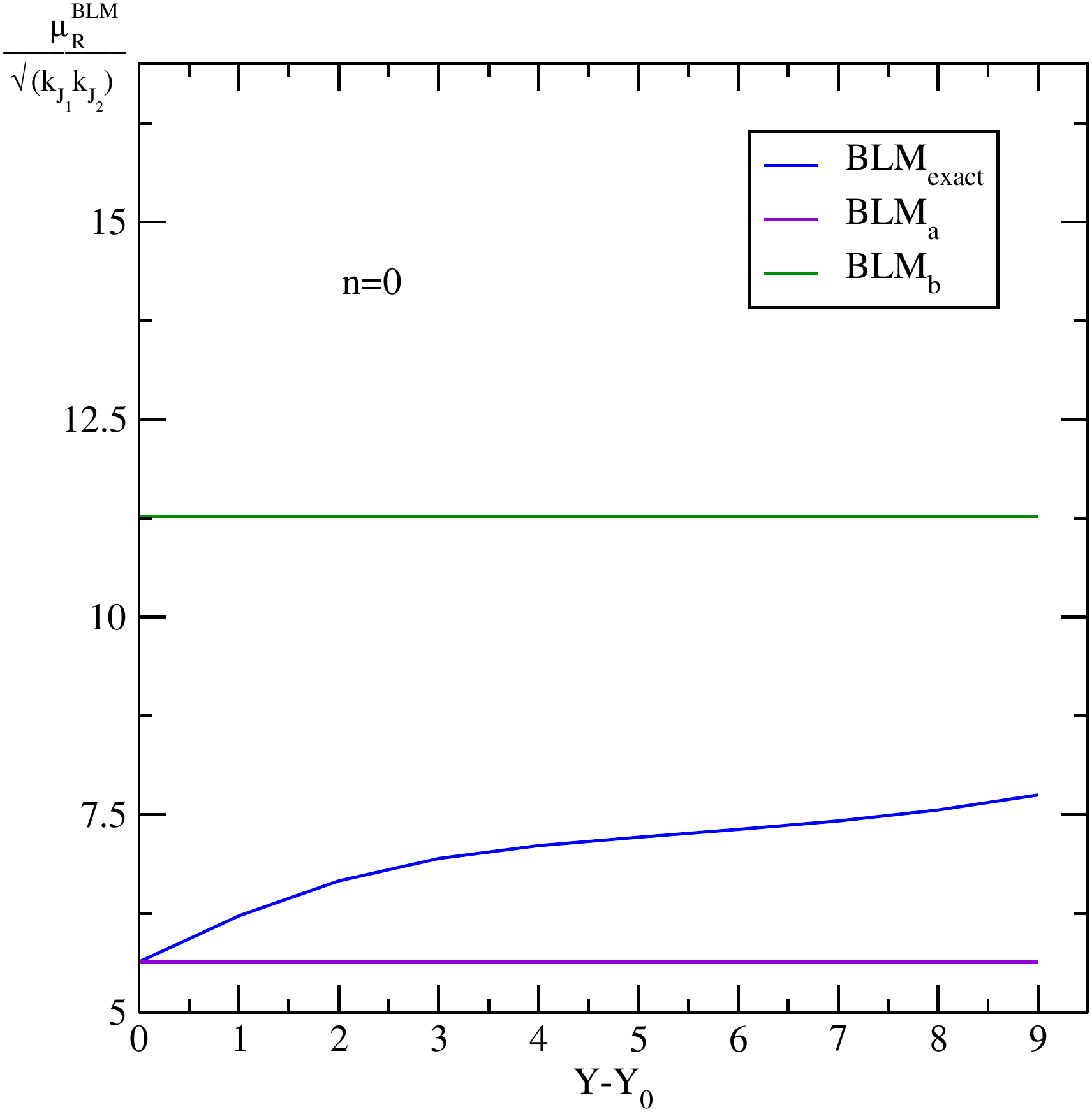}
\includegraphics[scale=0.45]{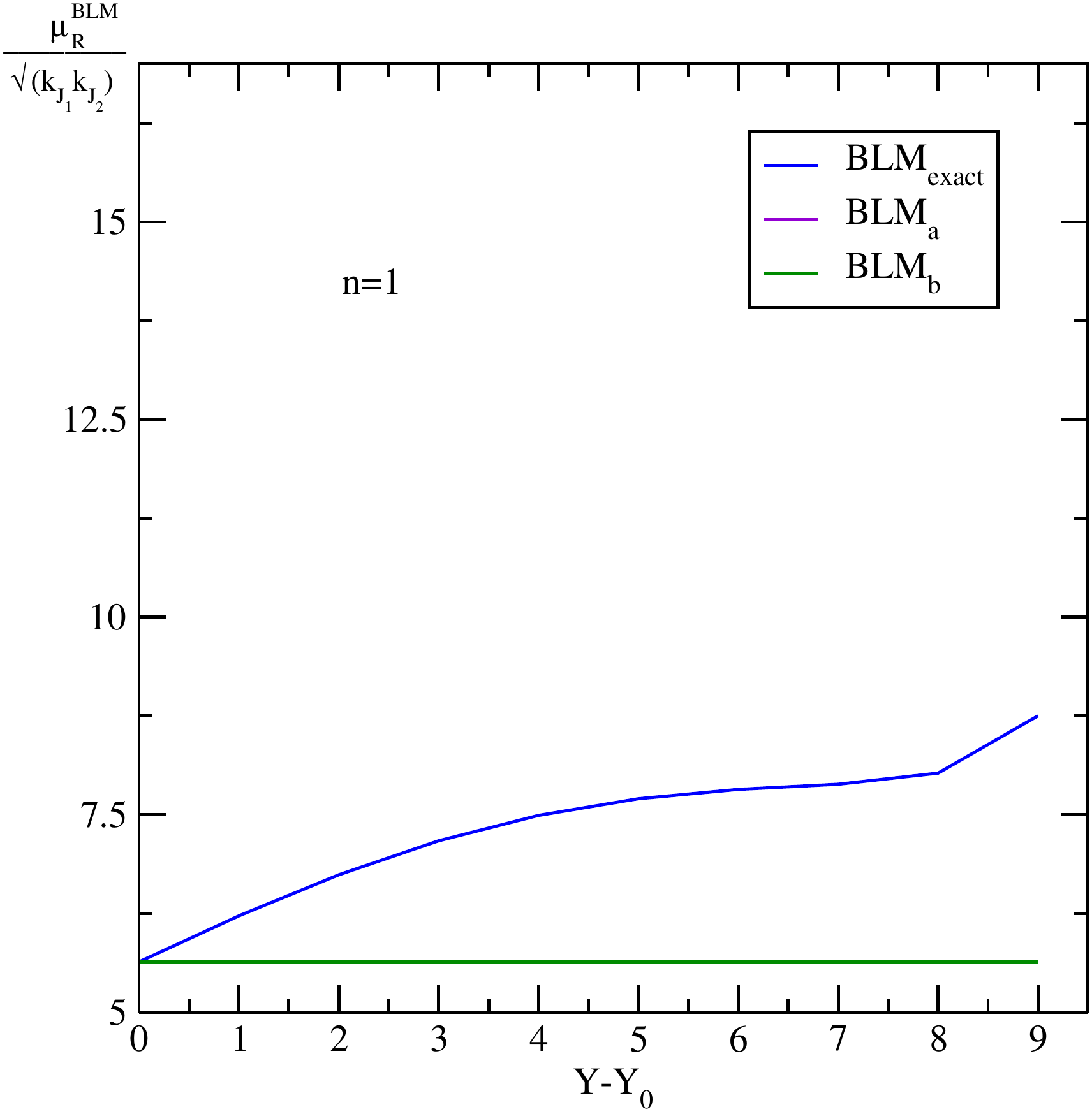}
\includegraphics[scale=0.45]{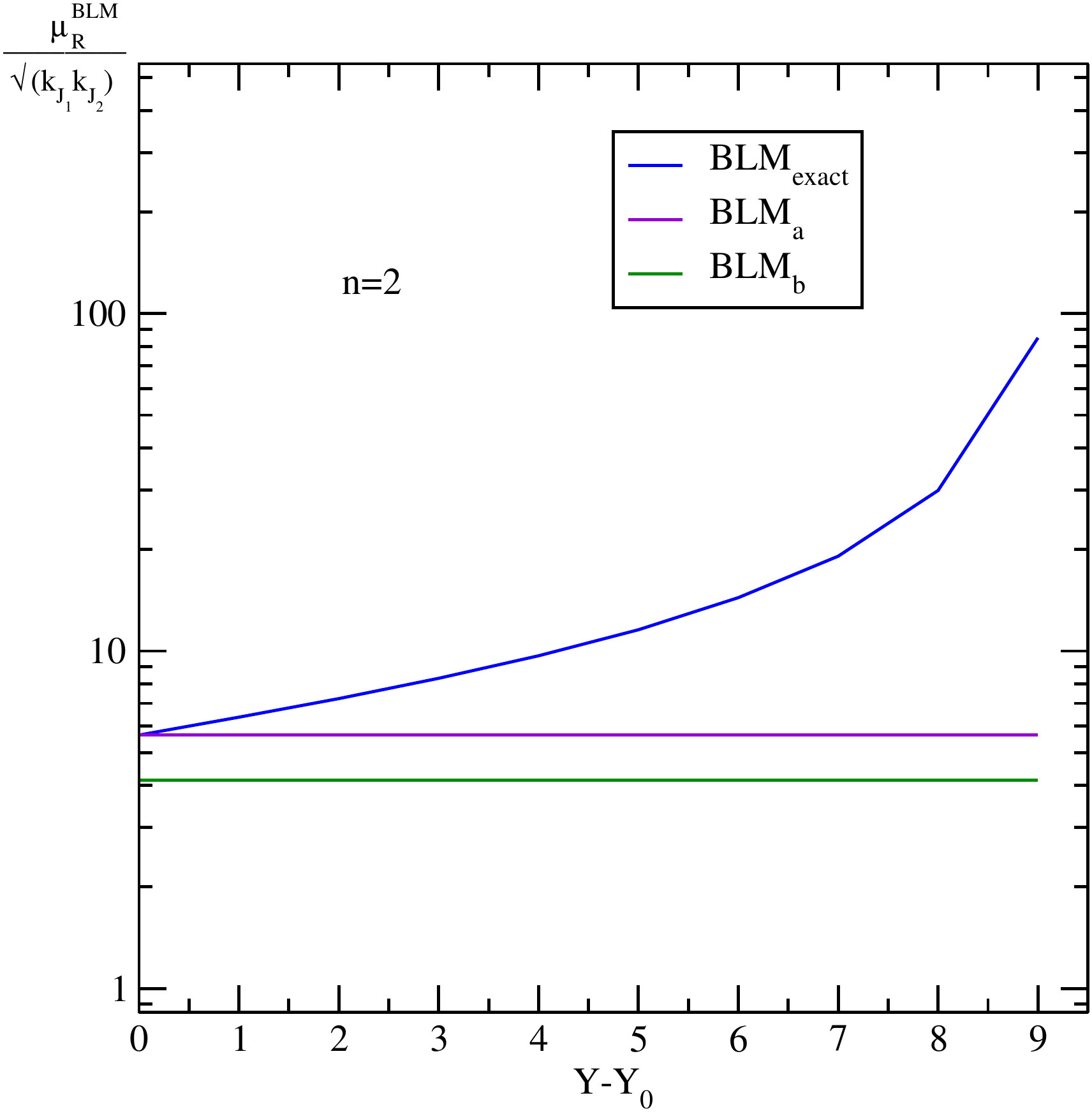}
\includegraphics[scale=0.45]{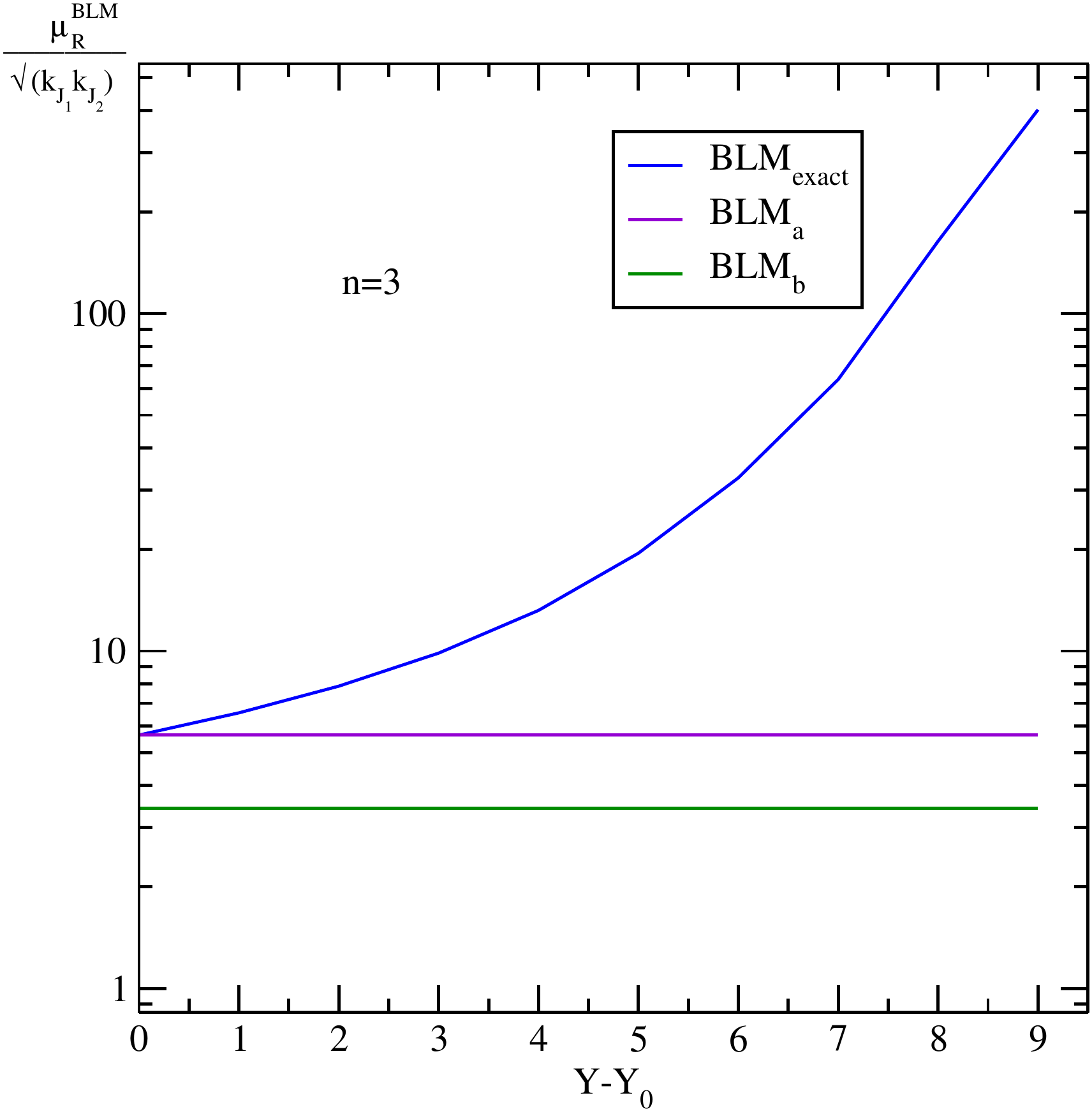}
\caption[]{BLM scales for Mueller-Navelet jets (see the text for details).}
\label{scalejet}
\end{figure}

\begin{figure}[t]
\centering
\includegraphics[scale=0.44]{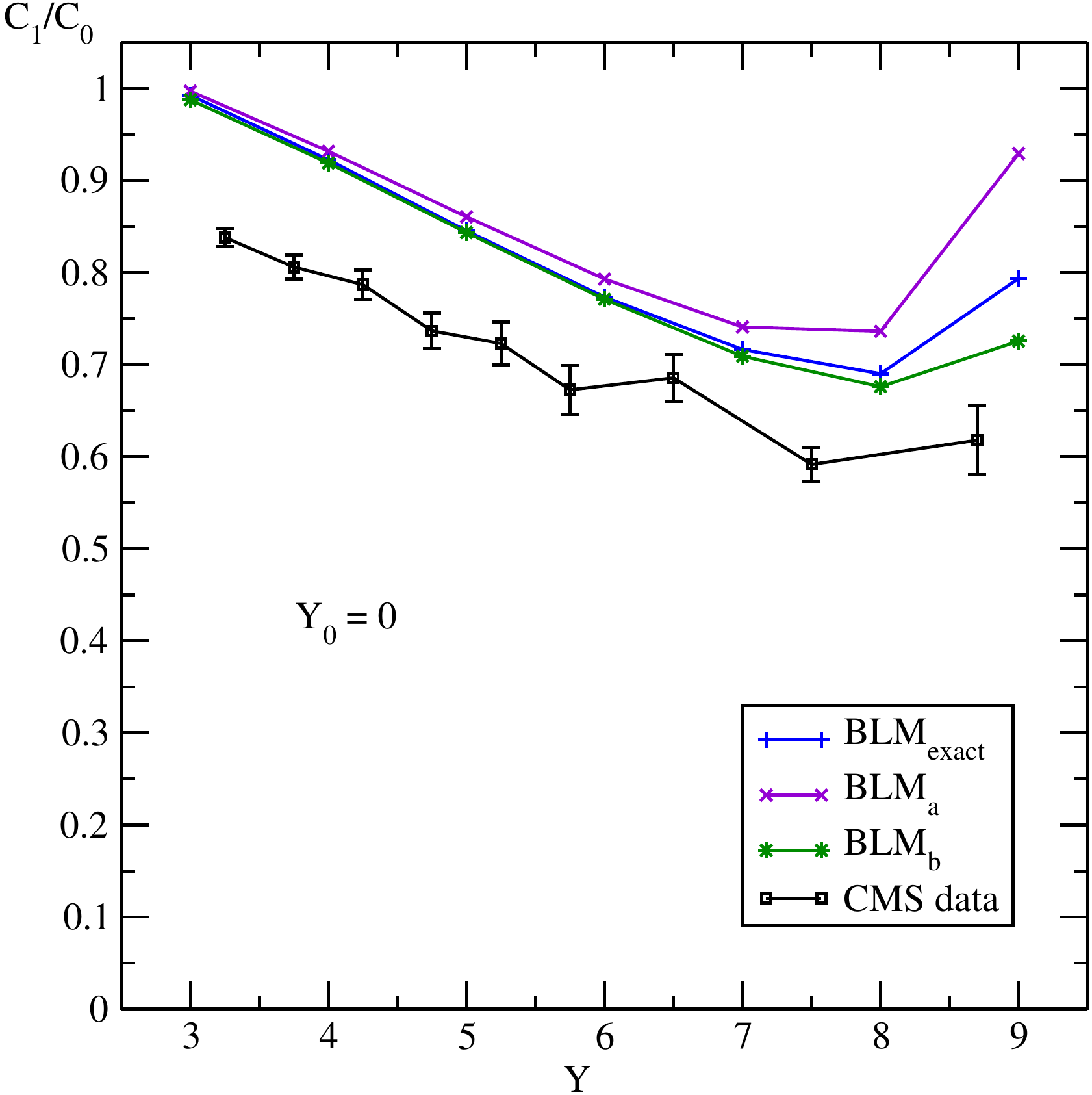}
\includegraphics[scale=0.44]{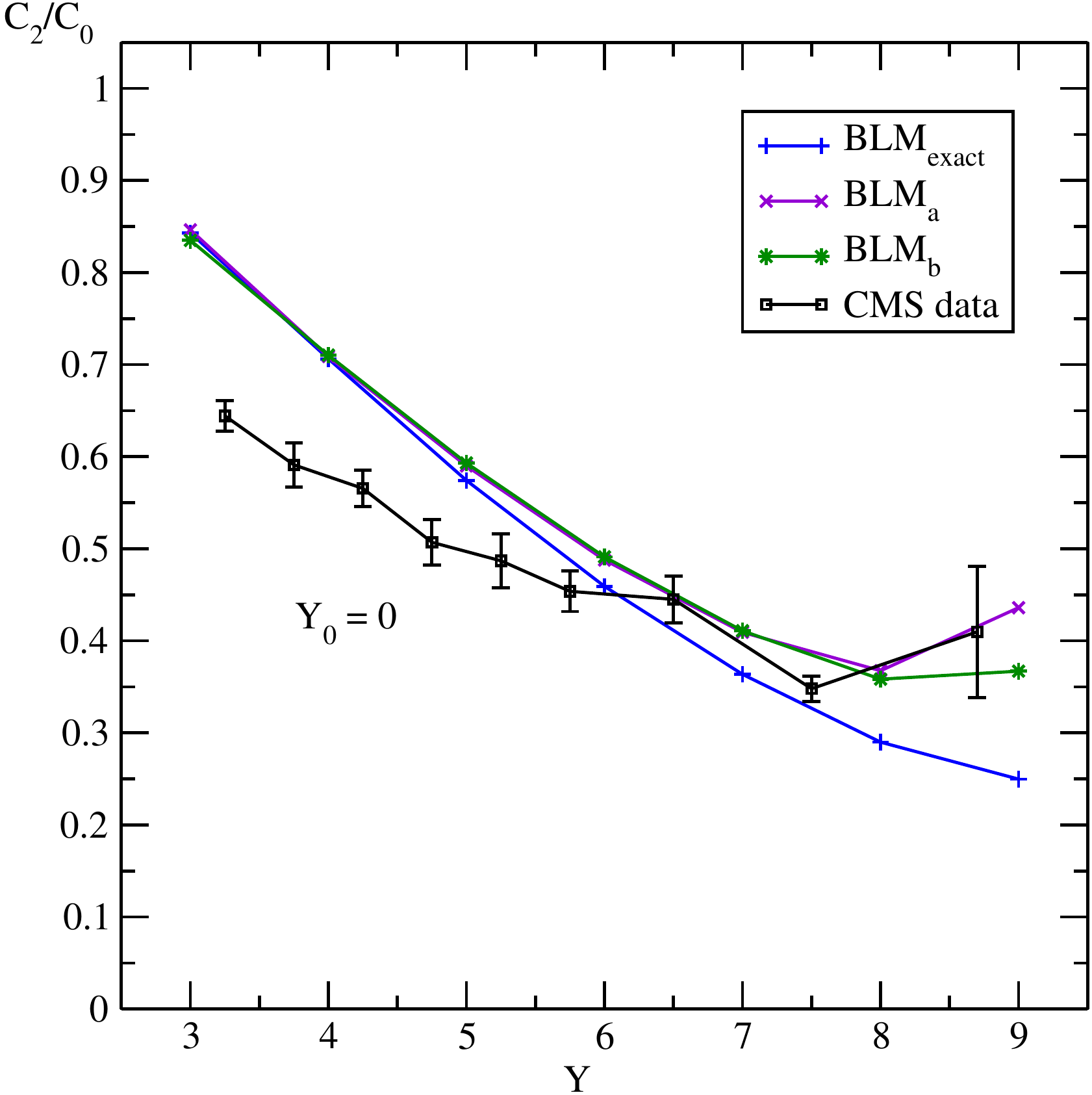}
\includegraphics[scale=0.44]{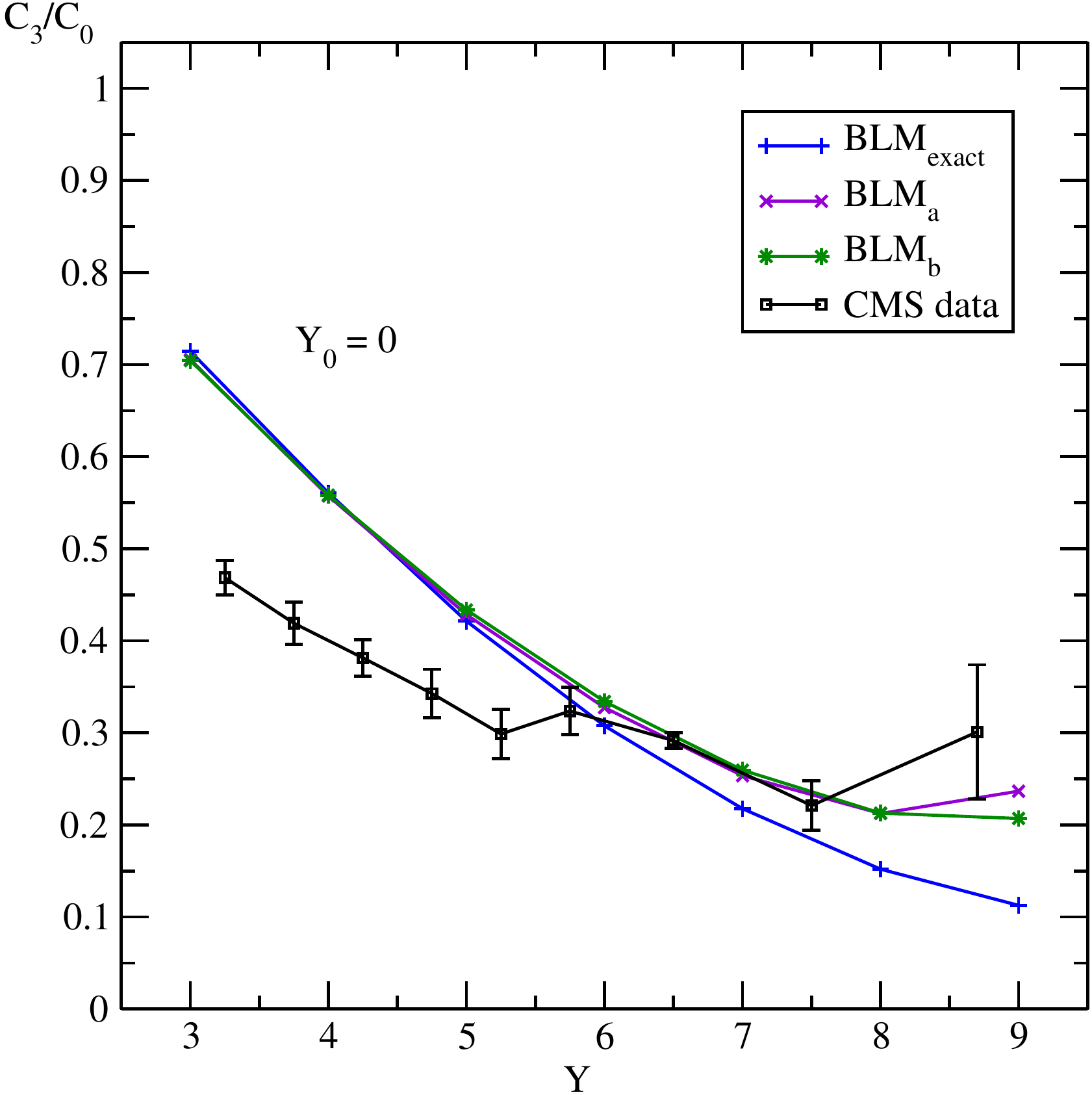}
\includegraphics[scale=0.44]{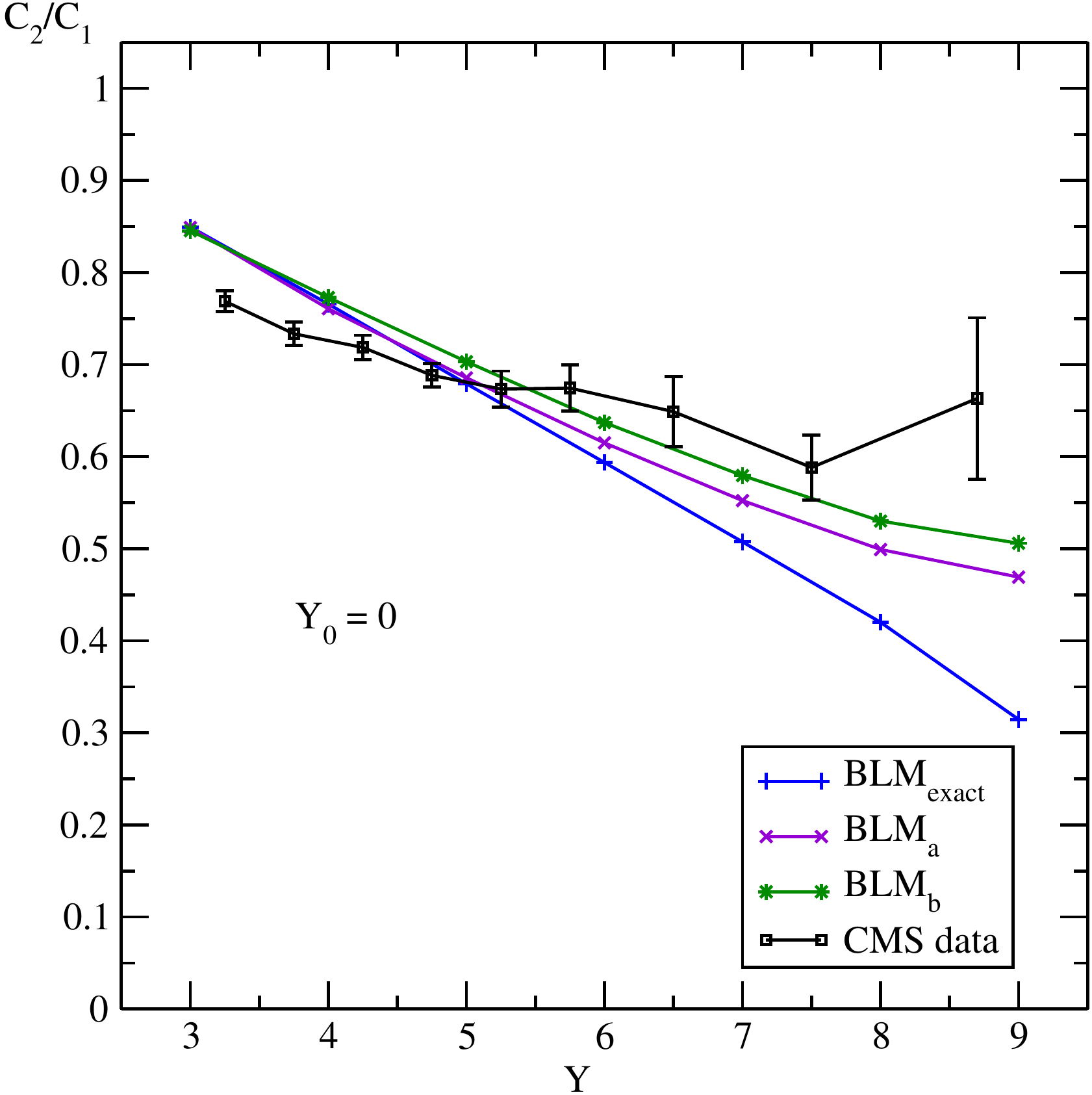}
\includegraphics[scale=0.44]{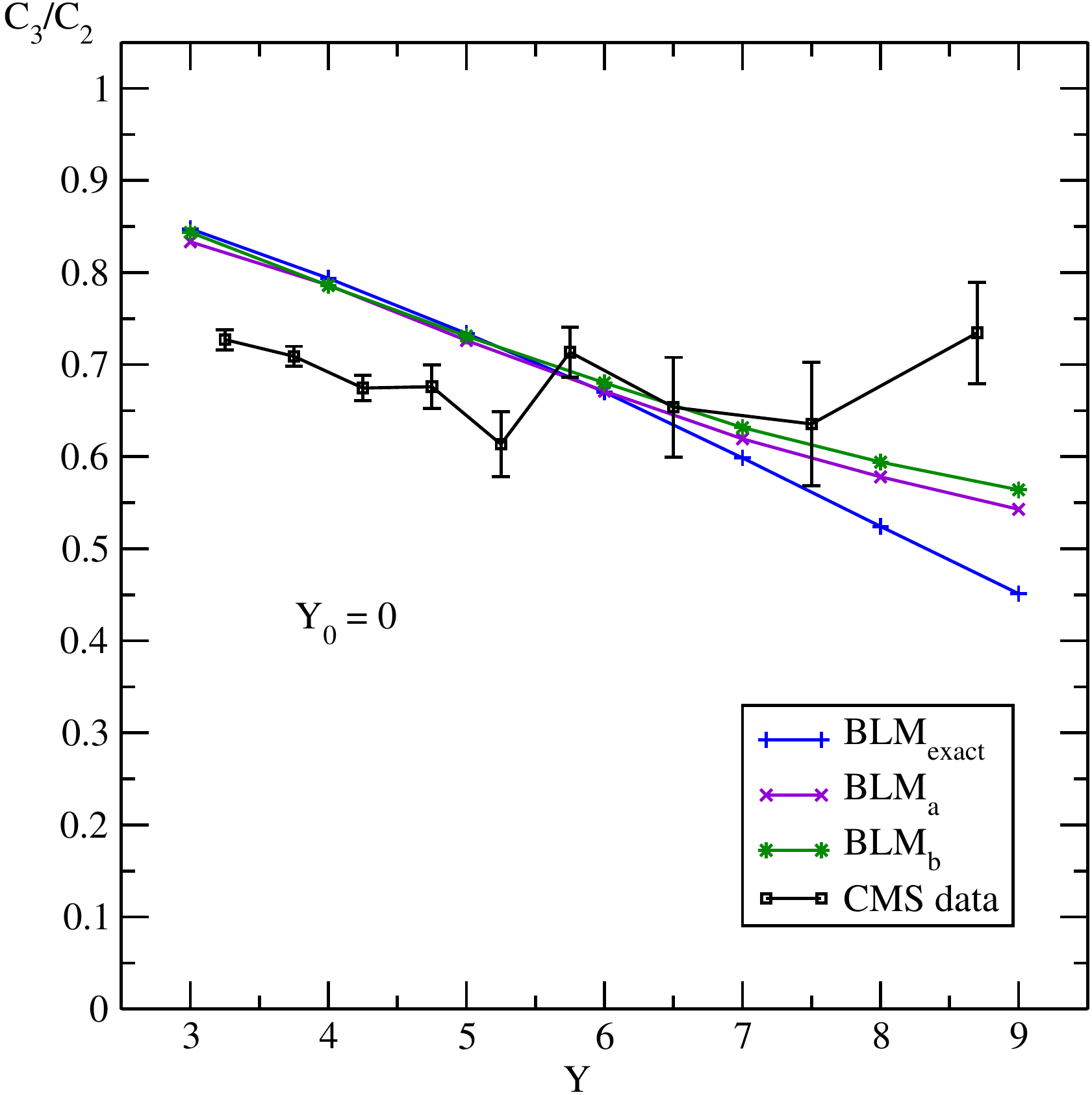}
\caption[]{Azimuthal decorrelations for Mueller-Navelet jets (see the text
for details).}
\label{figratiosY00}
\end{figure}

In this section we use the same kinematic settings as in~\cite{Caporale2014}
and present the BLM scale setting for the Mueller-Navelet jet production.
In particular, we consider the ratios $\mu_R^{\rm BLM}/\sqrt{k_{J_1}k_{J_2}}$
as functions of $Y-Y_0$ for $n$=0, 1, 2 and~3 and recall that, for this
process, the function $f(\nu)=0$ is zero. The results are shown in 
Fig.~\ref{scalejet} were the three lines, violet, green and blue,
denote the cases~$(a)$, $(b)$ and ``exact'', respectively. For this process 
it is not possible to consider the case~$(c)$ because the product of LO impact 
factors $c_1(n,\nu)c_2(n,\nu)$ is a function that does not decreases so fast, 
so that the $\nu$-interval needed for the integration includes the value 
$\overline\nu$, defined by Eq.~(\ref{barnu}), where the method is not
applicable.

Due to the integration over the jet variables $k_{J_{1,2}}$ and $y_{J_{1,2}}$, the 
derivation of the ``exact'' curve here is a little bit different from that 
of the other two processes. In this case, in order to get the 
ratios $\mu_R^{\rm BLM}/\sqrt{k_{J_1}k_{J_2}}$, we write 
$\mu_R=m_R\sqrt{k_{J_1}k_{J_2}}$ and we look for $m_R$ such that 
Eq.~(\ref{c_nnnbeta}) is satisfied. 

On the contrary, since cases~$(a)$ and $(b)$ are independent of the energy
of the process for $n=0$, the two curves, ``BLM$_a$'' and ``BLM$_b$'' are 
equivalent to those in Fig.~\ref{fig:photons}, since also in the present case
$f(\nu)=0$. 
Moreover, note that, for $n=1$, $\chi(n=1, \nu=0)=0$ and therefore 
in Fig.~\ref{scalejet} the curve ``BLM$_b$'' overlaps exactly the curve 
``BLM$_a$''.

In Fig.~\ref{figratiosY00} we present some ratios $C_m/C_n$ {\it versus} 
$Y$, where we make use of the scales shown in Fig.~\ref{scalejet}. In all
cases shown in Fig.~\ref{figratiosY00} the factorization scale $\mu_F$
entering the MSTW2008nlo~\cite{PDF} parton distribution functions
was chosen equal to the renormalization scale $\mu_R$ and the BFKL energy scale 
$Y_0$ was fixed at zero. One could look for optimal choices of the scale $Y_0$,
based on the PMS method, for instance, but this goes beyond the scope
of the present paper. 
The results shown in~\cite{Caporale2014} are a little bit 
different from those shown here, because there we transferred back all 
formulas to the $\overline{\rm MS}$ scheme. Moreover note that now we have an 
extra-curve (BLM$_{\rm exact}$) in which $\mu_R$ was obtained solving 
Eq.~(\ref{c_nnnbeta}).

\section{Summary}

In this paper we have focused on the BLM method to set the renormalization
scale in a ge\-ne\-ric se\-mi\-hard pro\-cess, as described in the NLA BFKL 
approach in the $(\nu,n)$-represen\-ta\-tion.
We found that the BLM scale setting procedure is well defined
in the context of semihard processes described by the BFKL approach within 
NLA accuracy.
The straightforward application of the BLM procedure leads
to a condition to be fulfilled, Eq.~(\ref{c_nnnbeta}), which defines the 
optimal renormalization scale depending on the specific process and on its 
energy. Our main observation here is that, due to the presence of 
$\beta_0$-terms in the next-to-leading expressions for the process dependent 
impact factors, the optimal renormalization scale is not universal, but 
turns out to depend both on the energy and on the type of process in question. 
The non-universality of the BLM scale setting in exclusive processes
was observed already in~\cite{Anikin:2004jb}.

Note that the above-mentioned $\sim \beta_0$-contributions to NLA impact 
factors are universally expressed in terms of the LO impact factors of the 
considered process, see our Eqs.~(\ref{beta-if}) and~(\ref{if2}).
Thus, they could be easily calculated for all processes, even in the case when 
the full expressions for the NLO corrections to the impact factors are not 
known. Such contributions must be taken into account in the implementation of 
BLM method to the description of cross sections of semihard processes, because 
{\it all} contributions to the cross section that are $\sim \beta_0$ must
vanish at the BLM scale. 

Such an ``exact'' implementation of the BLM method could be difficult since 
it calls for the solution of an integral equation,  Eq.~(\ref{c_nnnbeta}),  
for each value of the energy of the process. 
This equation can be solved, in general, only in numerical way. 
Therefore, we considered several approximated approaches to the BLM scale 
setting. One of them, the closest to the ``exact'' one and labeled $(c)$, 
consists in imposing the vanishing of the integrand appearing in the 
above-mentioned general condition and leads to an optimal BLM scale depending 
also on the $\nu$-variable. This approximated method has a validity domain in 
the $\nu$-space and can be applied only if the relevant range of the 
$\nu$-integration giving a physical observable falls inside this validity 
domain. Other approximated approaches, labeled $(a)$ and $(b)$, can be viewed 
as a sort of low- and high-energy approximation of the case~$(c)$ and of 
the ``exact'' determination.

We have compared these different approaches in the study of the total cross 
section and of other physical observables related with the forward amplitude
in processes such as the electroproduction of two light vector mesons,
the total cross section of two virtual photons and the production of 
Mueller-Navelet jets~\footnote{For all these cases the expression of the 
amplitude is known within the NLA as convolution of NLO impact factors with the 
NLA BFKL Green's function.}. Note that the formulas for the approximate
cases~$(a)$ and $(b)$ were already used by us, without derivation, 
in our recent papers~\cite{Ivanov:2014hpa,Caporale2014}. 
Here we presented in full details the implementation of the BLM method for 
arbitrary semihard processes, considering both its exact and approximate forms.

We could observe that, in general, the BLM scale setting in the cases~$(a)$ 
and $(b)$ provides with a range inside which lie the ``exact'' 
the case~$(c)$ determinations. This is not the case for the Mueller-Navelet
jet production where, as discussed in the text, due to some peculiarities
in the definition of the observables imposed by the experimental cuts, the 
natural ordering between the optimal scales in the cases~$(a)$, $(b)$ and
``exact'' is sometimes lost. It turns out, however, that azimuthal correlations
and ratios between them in the Mueller-Navelet case are less sensitive to
the different approaches to BLM scale setting than in the other two processes
considered in this work.

Note that previous applications of the BLM method to the description of 
$\gamma^*\gamma^*$ total cross 
sections~\cite{Brodsky:1996sg,Brodsky:1997sd,Brodsky:2002ka,Zheng:2013uja} 
relied on the use of LO expressions for the photon impact factors. 
In~\cite{Brodsky:1996sg,Brodsky:1997sd} the $\gamma^*\gamma^*$ total cross 
section was considered in LLA BFKL, since the NLO corrections to the BFKL kernel
were not yet known. However, in~\cite{Brodsky:1996sg,Brodsky:1997sd} the 
$\beta_0$-part of the first correction to the Born amplitude ({\it i.e.} the 
$t$-channel two-gluon exchange) was considered in order to establish the 
renormalization scale. Such approach to the scale setting is closely related 
to our case~$(a)$ (scale fixed from the correction to the impact factor). 
Indeed, considering the expansion of the BFKL amplitude~(\ref{ampl-ff}), one 
can see that the first, $\sim \alpha_s$, correction to the Born amplitudes 
originates entirely from NLO parts of the impact factors.  
Comparing Eq.~(5.5) in~\cite{Brodsky:1997sd} with our Eq.~(\ref{casea}) for
$f(\nu)=0$, as appropriate for the $\gamma^*\gamma^*$ process, one can 
see that they agree except for the term that, in our approach, derived from
the change to the MOM scheme. One can therefore refer 
to~\cite{Brodsky:1996sg,Brodsky:1997sd} as to the first (approximate) 
application of the BLM scale setting to a BFKL calculation. 
In~\cite{Brodsky:2002ka,Zheng:2013uja} the $\gamma^*\gamma^*$ total cross was 
considered using full NLA BFKL kernel, but with LO approximation for the 
photon impact factor. In respect with the BLM scale setting, such approach is 
equivalent to our approximate case~$(b)$.

In~\cite{Ducloue2014} the BLM method was applied to Mueller-Navelet 
jet production: although the full NLO expression for the jet impact factor 
was used, the above-discussed effect of $\beta_0$-contributions to NLO 
jet impact factors on the choice of the BLM scale was overlooked. Therefore 
in~\cite{Ducloue2014} the value of the BLM scale which was obtained 
is similar to the one used in~\cite{Brodsky:2002ka,Zheng:2013uja} and, as 
such, coincides with our approximate case~$(b)$. 
Our results presented in Fig.~\ref{figratiosY00} allow to assess the inaccuracy 
in BLM predictions for different Mueller-Navelet jet observables related with 
approximated approaches to the BLM scale setting.  

In conclusion, the BLM method for scale setting, which was proposed 
more than three decades ago on a strong physical basis, remains a
fundamental tool for perturbative calculations and has lead to many successful
comparisons between theoretical predictions and experimental data.
In this paper we have provided with the general paradigm
for its systematic application to an important class of processes, {\it i.e.} 
semihard processes within the NLA BFKL approach, thus filling some gaps
left open by previous approximated or incomplete approaches. We believe that
this will increase the future significance of the method.

\section*{Acknowledgements}

D.I. thanks the Dipartimento di Fisica dell'U\-ni\-ver\-si\-t\`a della Calabria
and the Istituto Nazio\-na\-le di Fisica Nucleare (INFN), Gruppo collegato di
Cosenza, for the warm hospitality and the financial support. The work
of D.I. was also supported in part by the Russian Foundation for Basic Research
via grant RFBR-13-02-00695-a.

The work of B.M. was supported by the European Commission, European Social Fund 
and Calabria Region, that disclaim any liability for the use that can be done 
of the information provided in this paper.
B.M. thanks the Sobolev Institute of Mathematics of Novosibirsk for warm 
hospitality during the preparation of this work.

\end{document}